\documentclass[12pt,english]{article}
\usepackage[T1]{fontenc}
\usepackage[utf8]{inputenc}
\usepackage[pdftex]{geometry}
\usepackage{bm}
\usepackage{amsmath}
\usepackage{amssymb}
\usepackage[pdftex]{graphicx}
\usepackage{setspace}
\makeatletter

\providecommand{\tabularnewline}{\\}

\numberwithin{equation}{section}
\newcommand{\lyxaddress}[1]{
	\par {\raggedright #1
	\vspace{1.4em}
	\noindent\par}
}

\allowdisplaybreaks[1]

\usepackage{braketmod}
\usepackage{color}

\usepackage[pdftex]{hyperref}
\hypersetup{
colorlinks,
linktoc=page,
citecolor=blue,
linkcolor=blue,
urlcolor=blue
}

\date{}

\makeatother

\usepackage{babel}
\usepackage[style=phys,biblabel=brackets, chaptertitle=false, pageranges=false, eprint=true]{biblatex}
\begin{document}
\begin{flushright}
{\small{}YITP-21-153}{\small\par}
\par\end{flushright}

\noindent\begin{minipage}[t]{1\columnwidth}%
\title{\textbf{3d dualities with decoupled sectors}\\
\textbf{and brane transitions}}
\author{Naotaka Kubo\,\footnotemark}
\maketitle

\lyxaddress{\begin{center}
\vspace{-18bp}
$^{*}$\,\textit{Center for Gravitational Physics, Yukawa Institute for Theoretical Physics,}\\
\textit{Kyoto University, Sakyo-ku, Kyoto 606-8502, Japan}\vspace{-10bp}
\par\end{center}}
\begin{abstract}
We find exact relations among the sphere partition functions of three-dimensional $\mathcal{N}=4$ superconformal Chern-Simons theories with circular quiver diagrams. These relations suggest new dualities in gauge theories which are the products of circular quiver gauge theories and decoupled linear quiver gauge theories. The dualities can be interpreted as simple brane transitions in the Hanany-Witten brane construction in type IIB string theory. Interestingly, the brane transitions cannot be generated by the Hanany-Witten transition or the ${\rm SL}\left(2,\mathbb{Z}\right)$ transformation. Our results can be regarded as generalization and clarification of dualities implied from Weyl group symmetries of quantum curves. To obtain the exact results, we employ the supersymmetric localization technique and the Fermi gas approach.
\end{abstract}
\end{minipage}

\renewcommand{\thefootnote}{\fnsymbol{footnote}}
\footnotetext[1]{\textsf{naotaka.kubo@yukawa.kyoto-u.ac.jp}}
\renewcommand{\thefootnote}{\arabic{footnote}}
%\today

\newpage{}

\global\long\def\lp{\text{1-loop}}%

\global\long\def\cs{\text{CS}}%

\global\long\def\vector{\text{vec}}%

\global\long\def\matter{\text{mat}}%

\global\long\def\cstot{\text{CSs}}%

\global\long\def\bifund{\text{vec/mat}}%

\global\long\def\fermi{\text{FG}}%

\global\long\def\bra#1{\Bra{#1}}%

\global\long\def\bbra#1{\Bbra{#1}}%

\global\long\def\ket#1{\Ket{#1}}%

\global\long\def\kket#1{\Kket{#1}}%

\global\long\def\braket#1{\Braket{#1}}%

\global\long\def\bbraket#1{\Bbraket{#1}}%

\global\long\def\brakket#1{\Brakket{#1}}%

\global\long\def\bbrakket#1{\Bbrakket{#1}}%

\tableofcontents{}

\section{Introduction}

The brane dynamics has turned out to provide an intuitive and unified understanding of many types of dualities in gauge theories. For instance, in the Hanany-Witten brane construction, the Hanany-Witten transition and the ${\rm SL}\left(2,\mathbb{Z}\right)$ transformation provide many dualities such as Aharony duality, Giveon-Kutasov duality and Mirror symmetry of three-dimensional supersymmetric quiver gauge theories (see \cite{Hanany:1996ie,deBoer:1996ck,Giveon:2008zn,Amariti:2015yea} for example). These dualities claim that the dual theories flow to the same interacting CFT in the IR fixed point.

In this paper, we study the worldvolume theories of D3-branes in the Hanany-Witten setup with NS5-branes and $\left(1,k\right)$5-branes. The ABJ(M) theory \cite{Aharony:2008ug,Hosomichi:2008jb,Aharony:2008gk}, which is $\mathcal{N}=6$ superconformal Chern-Simons theory with ${\rm U}\left(N_{1}\right)_{k}\times{\rm U}\left(N_{2}\right)_{-k}$ gauge group, is a famous and important example. The subscripts denote the Chern-Simons levels. The corresponding brane configuration consists of one NS5-brane, one $\left(1,k\right)$5-brane and D3-branes. The number of the 5-branes in this setup can be generalized to an arbitrary number. In this case, the worldvolume theory is an $\mathcal{N}=4$ superconformal Chern-Simons theory with a linear or circular quiver diagram \cite{Gaiotto:2008sd,Hosomichi:2008jd,Imamura:2008dt}.

The sphere partition function can be used to test dualities in the IR fixed point since the partition function is invariant under the RG flow. Thanks to the supersymmetric localization technique \cite{Pestun:2007rz,Kapustin:2009kz}, the path integral computing the partition function is reduced to finite-dimensional integral, which we call a matrix model. However, the matrix model is still difficult to study, especially in the small Chern-Simons level region, which is called the M-theory region. For a class of $\mathcal{N}\geq3$ Chern-Simons theories, the Fermi gas approach \cite{Marino:2011eh} was a breakthrough for this problem. The idea of this approach is to rewrite the matrix model to the partition function of an ideal Fermi gas. Even though the Fermi gas formalism was developed for studying the M-theory region, this formalism has also yielded many other useful results. One important aspect is that, similar to the brane picture, the density matrix of the Fermi gas has been turned out to provide a unified understanding of mirror symmetries of $\mathcal{N}\geq4$ superconformal Chern-Simons theories \cite{Drukker:2015awa,Assel:2015hsa,Honda:2015rbb}. Because the inverse of density matrix of the Fermi gas associated to the gauge theory is known to become a quantum curve \cite{Marino:2011eh,Marino:2015ixa,Kashaev:2015wia,Hatsuda:2016uqa,Kubo:2019ejc,Kubo:2020qed}, it is natural to expect that the quantum curve also provides a new perspective for understanding the dualities in the $\mathcal{N}=4$ superconformal Chern-Simons theories.

Indeed, the study along this direction has been performed by studying the symmetry of the quantum curve. It was found that the genus one quantum curve enjoys Weyl group symmetry as well as the classical curve \cite{Kubo:2018cqw,Furukawa:2020cjp,Moriyama:2020lyk}. The information of the Weyl group symmetry implies the equality of the partition functions, and thus it can be pulled back to the information of the dualities in the gauge theories. Because the gauge theory which we consider now is the worldvolume theory of a brane configuration, the dualities imply brane transitions. Surprisingly, the Weyl group symmetry includes not only the Hanany-Witten transition and the ${\rm SL}\left(2,\mathbb{Z}\right)$ transformation, but also new brane transitions \cite{Kubo:2019ejc,Kubo:2020qed,Furukawa:2020cjp}. Recently, motivated by these dualities, local brane transitions were conjectured in \cite{Furukawa:2020cjp}.

However, the above consideration has some problems. First, the implication for dualities among the gauge theories from the quantum curve is restricted to special cases where the corresponding brane configuration consists of just two NS5-branes and a small number of the $\left(1,k\right)$5-branes. This is because there are only a small number of known relations between the matrix models and the quantum curves due to the lack of computational technique. Furthermore, the symmetry of the quantum curve has been studied only for genus one cases. Second, the quantum curve does not take into account the FI term, which is a natural deformation of the gauge theory including ${\rm U}\left(1\right)$ factors. Moreover, the FI term can be naturally introduced by shifting 5-branes to transverse directions to the D3-branes. Therefore, it is essential to include the FI terms for studying the brane dynamics. Third, in some cases, though the relations between the matrix models and the quantum curves were conjectured, these relations are still not proven. Fourth, the quantum curve cannot provide a complete analysis of the sphere partition functions because the quantum curves do not have all of the information about the corresponding matrix model.

In this paper, we resolve these problems. We focus on the matrix model, and we show exact relations directly without using the quantum curve. In this strategy, there is no longer any reason to limit ourselves to the matrix model which has an explicit relation to the quantum curve. Indeed, we deal with brane configurations including an arbitrary number of 5-branes and thus we obtain new functional relations beyond the results from the quantum curve. We improve the Fermi gas approach so that we can consider the deformation with the FI terms and we can prove the identity in the matrix model. Because we study the matrix model itself, our result is also sensitive to the factor in the matrix model to which the quantum curve cannot access. As a result, we find that we should add a decoupled linear quiver gauge theory to make the dualities correct.

Before closing this section, we present a simple example of the dualities we will investigate.
\begin{figure}
\begin{centering}
\includegraphics[scale=0.4]{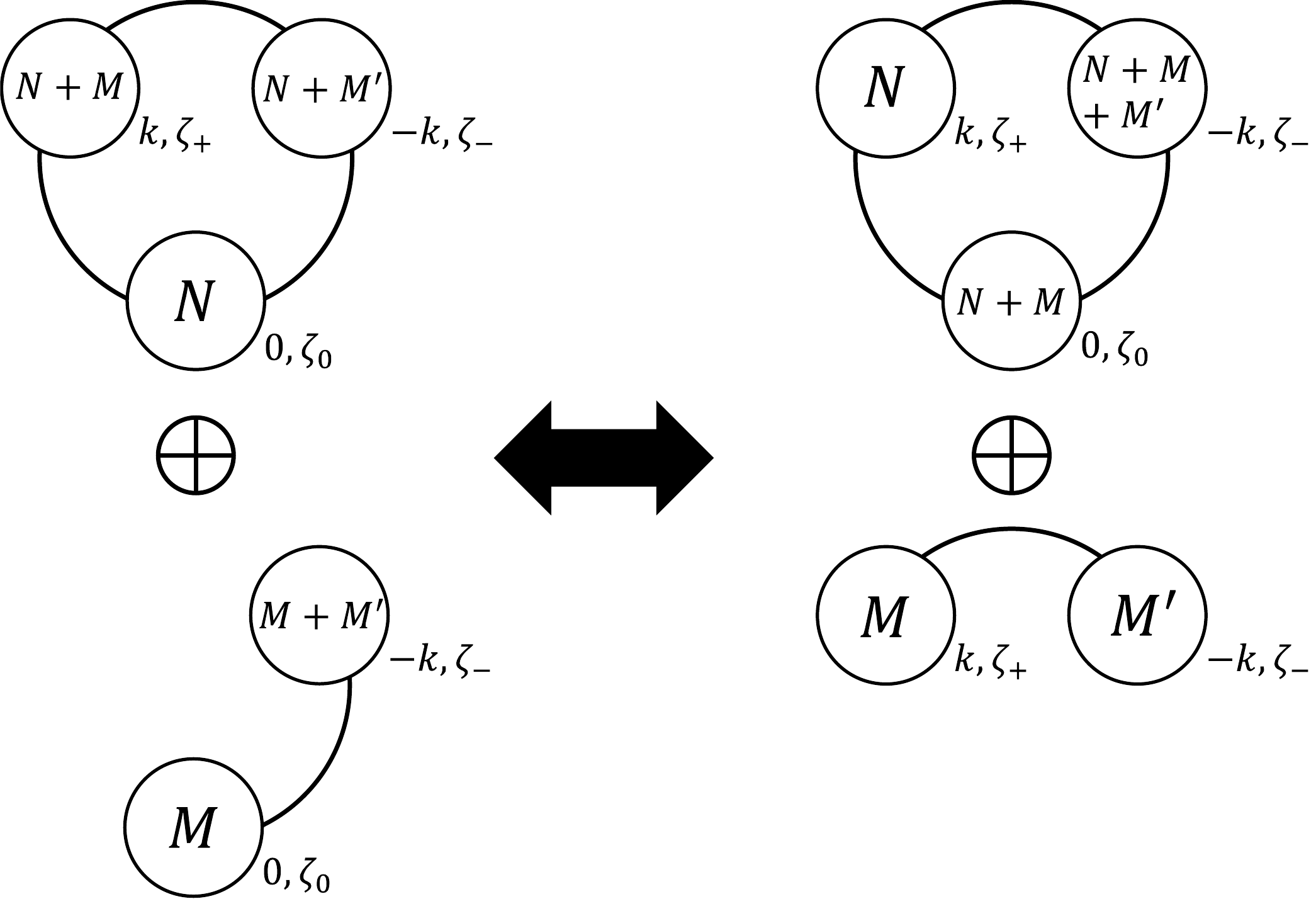}
\par\end{centering}
\caption{An example of the dualities. Each circle with a number $L$ at the center and subscripts $k,\zeta$ indicates an $\mathcal{N}=4$ ${\rm U}\left(L\right)$ vector multiplet whose Chern-Simons level and FI parameter are $k$ and $\zeta$, respectively. Each line indicates a bi-fundamental hypermultiplet. The theory on each side is the products of the circular quiver theory and the decoupled linear quiver theory. \label{fig:21Dual-Quiver}}
\end{figure}
Figure \ref{fig:21Dual-Quiver} shows the duality. The theory at the upper left is the $\mathcal{N}=4$ superconformal Chern-Simons theory with the gauge group
\begin{equation}
\left[\mathrm{U}\left(N+M\right)_{k}\times\mathrm{U}\left(N+M'\right)_{-k}\times\mathrm{U}\left(N\right)_{0}\right]^{\mathrm{C}}.
\end{equation}
The symbol ${\rm C}$ stresses that the quiver diagram is circular. The FI parameters should satisfy $\zeta_{+}+\zeta_{-}+\zeta_{0}=0$. We find that this theory is dual to the theory at the upper right of figure \ref{fig:21Dual-Quiver}: the $\mathcal{N}=4$ superconformal Chern-Simons theory with the gauge group
\begin{equation}
\left[\mathrm{U}\left(N\right)_{k}\times\mathrm{U}\left(N+M+M'\right)_{-k}\times\mathrm{U}\left(N+M\right)_{0}\right]^{\mathrm{C}}.
\end{equation}
However, to complete the duality, we have to add decoupled sectors for both sides. The decoupled theory which appear in the opposite side can be read off by substituting all of ranks by the lowest rank. In this case, we set $N=0$. As a consequence, the decoupled theory is always described by a linear quiver diagram. Therefore, the precise duality is
\begin{align}
 & \left[\mathrm{U}\left(N+M\right)_{k}\times\mathrm{U}\left(N+M'\right)_{-k}\times\mathrm{U}\left(N\right)_{0}\right]^{\mathrm{C}}\oplus\mathrm{U}\left(M+M'\right)_{-k}\times\mathrm{U}\left(M\right)_{0}\nonumber \\
 & \leftrightarrow\left[\mathrm{U}\left(N\right)_{k}\times\mathrm{U}\left(N+M+M'\right)_{-k}\times\mathrm{U}\left(N+M\right)_{0}\right]^{\mathrm{C}}\oplus\mathrm{U}\left(M\right)_{k}\times\mathrm{U}\left(M'\right)_{-k},\label{eq:21Dual-Ex}
\end{align}
as in figure \ref{fig:21Dual-Quiver}. The symbol $\oplus$ indicates the product of two decoupled theories. The rule of the duality can be easily understood by using the brane configuration. Therefore, in this paper, we first explain the dualities in terms of the brane configuration, and then we explain them in terms of the gauge theory and the matrix model.

This paper is organized as follows. In section \ref{sec:BC-MM-QC}, we review the brane configuration, the matrix model and the quantum curve. In section \ref{sec:2q-model}, we present our result as the brane transitions and the identities in the matrix model when the brane configuration includes two NS5-branes. We also show that the relations suggest new dualities in $\mathcal{N}=4$ superconformal Chern-Simons theories which are the products of circular quiver theories and decoupled linear quiver theories. In section \ref{sec:pq-model}, we generalize our result in the previous section to brane configurations including any even number of NS5-branes. In section \ref{sec:Res-QC}, we explain the relation between our results and the implications from the quantum curve. Finally, in section \ref{sec:Discussion}, we summarize our results, discuss implications of our results to the topics related to the quantum curve and give future directions. In appendix \ref{sec:FGF_loc}, we present the Fermi gas formalism for a local theory in a way that suits our purposes. We also show the physical interpretations of factors which appear in the Fermi gas formalism. In appendix \ref{sec:HW}, we review the Hanany-Witten transition and how the transition can be seen in the matrix model.

\section{Setup and implications from quantum curve\label{sec:BC-MM-QC}}

In this section, we describe the setup and review relations between brane configurations, matrix models and quantum curves. We also review symmetries of the quantum curves and their implications to the brane configurations.

\subsection{Brane configuration\label{subsec:BC}}

In this paper, we focus on the type of the brane configuration which was first introduced by Hanany and Witten \cite{Hanany:1996ie}. This brane configuration consists of NS5-branes, D5-branes and D3-branes in type IIB string theory. Later, the type of 5-brane was generalized to $\left(p,q\right)$5-brane, which is the bound state of $p$ NS5-branes and $q$ D5-branes. In this case, generally, the worldvolume theory of the D3-branes in the IR fixed point becomes superconformal Chern-Simons theory \cite{Kitao:1998mf}.

We further focus on the brane configuration in which the worldvolume theory of D3-branes possesses at least $\mathcal{N}=4$ supersymmetry \cite{Gaiotto:2008sd,Hosomichi:2008jd,Aharony:2008ug,Imamura:2008dt}. The brane configuration consists of D3-branes, NS5-branes and $\left(1,k\right)$-5-branes. We assume $k>0$ for simplicity. The D3-branes are extended on 0126 direction with 6 direction compactified. The NS5-branes are extended on 012345 direction, while the $\left(1,k\right)$5-branes are extended on 012$\left[3,7\right]_{\theta}$$\left[4,8\right]_{\theta}$$\left[5,9\right]_{\theta}$ direction.
\begin{table}
\begin{centering}
\begin{tabular}{|c||c|c|c|c|c|c|c|c|c|c|}
\hline 
 & 0 & 1 & 2 & 3 & 4 & 5 & 6 ($S^{1}$) & 7 & 8 & 9\tabularnewline
\hline 
\hline 
D3-branes & $\circ$ & $\circ$ & $\circ$ &  &  &  & $\circ$ &  &  & \tabularnewline
\hline 
NS5-branes & $\circ$ & $\circ$ & $\circ$ & $\circ$ & $\circ$ & $\circ$ &  &  &  & \tabularnewline
\hline 
$\left(1,k\right)$5-branes & $\circ$ & $\circ$ & $\circ$ & $\slash_{37}$ & $\slash_{48}$ & $\slash_{59}$ &  & $\slash_{37}$ & $\slash_{48}$ & $\slash_{59}$\tabularnewline
\hline 
\end{tabular}
\par\end{centering}
\caption{The brane configuration which we study in this paper. $\slash_{ij}$ means that the brane is rotated in $\left(i,j\right)$ plane with appropriate angle so that the $\mathcal{N}=4$ supersymmetry is preserved.\label{tab:BC}}

\end{table}
 We summarize the directions in table \ref{tab:BC}. The 5-branes are separated on 6 direction. We can also move the position of the 5-branes to the direction transverse to the D3-branes. There are two types of D3-branes. The first type is the D3-branes which are going all the way around the circle. We call them regular D3-branes. The second type is the D3-branes which are not going all the way around the circle of the 6 direction.. We call them fractional D3-branes.

\begin{table}
\begin{centering}
\begin{tabular}{|c|c|}
\hline 
Elements of brane configuration & Pictorial notation\tabularnewline
\hline 
\hline 
NS5-brane with shift parameter $\zeta$ & $\bullet_{\zeta}$ (or $+1$)\tabularnewline
\hline 
$\left(1,k\right)$5-brane with shift parameter $\zeta$ & $\circ_{\zeta}$ (or $-1$)\tabularnewline
\hline 
The number of D3-branes & $N_{a}$\tabularnewline
\hline 
\end{tabular}
\par\end{centering}
\caption{Notation for the brane configuration.\label{tab:BCNotation}}
\end{table}
In this paper, we present the brane configuration by enumerating $\bullet_{\zeta}$ and $\circ_{\zeta}$ and putting $N_{a}$ between them. Here, $\bullet$ and $\circ$ denote an NS5-brane and a $\left(1,k\right)$5-brane, respectively, and the subscript $\zeta\in\mathbb{R}$ denotes the amount of shift of 5-branes to the transverse space.
\begin{figure}
\begin{centering}
\includegraphics[scale=0.7]{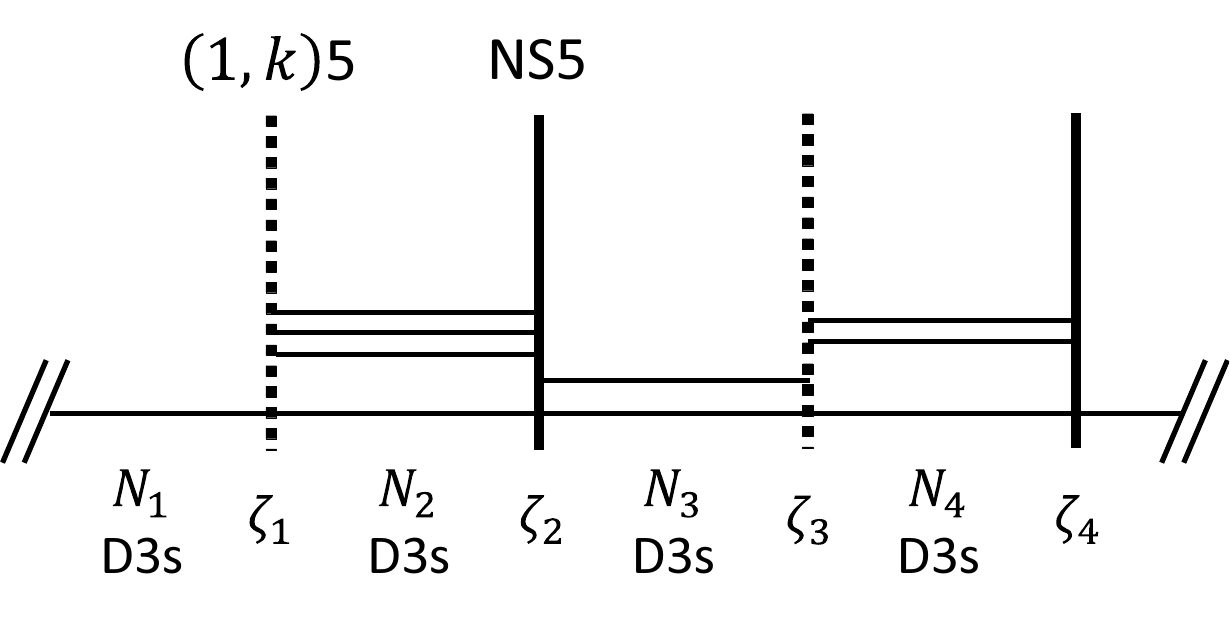}
\par\end{centering}
\caption{The brane configuration in \eqref{eq:BC-Ex1} when $N_{1}=1$, $N_{2}=4$, $N_{3}=2$ and $N_{4}=3$ The vertical solid lines denote the NS5-branes, the vertical dotted line denote the $\left(1,k\right)$5-branes and the horizontal lines denote the D3-branes. The double diagonal lines mean periodicity. \label{fig:BC-ex}}
\end{figure}
For example,
\begin{equation}
\left\langle N_{1}\circ_{\zeta_{1}}N_{2}\bullet_{\zeta_{2}}N_{3}\circ_{\zeta_{3}}N_{4}\bullet_{\zeta_{4}}\right\rangle ^{\mathrm{P}},\label{eq:BC-Ex1}
\end{equation}
denotes the brane configuration in figure \ref{fig:BC-ex}. $\mathrm{P}$ means that $6$ direction is periodic. When we focus on the part of the brane configuration, we remove $\mathrm{P}$ as
\begin{equation}
\left\langle N_{2}\bullet_{\zeta_{2}}N_{3}\circ_{\zeta_{3}}N_{4}\right\rangle .
\end{equation}
This is the brane configuration appearing in the center of \eqref{eq:BC-Ex1}. We call part of brane configuration local brane configuration. We also assign $s=+1$ and $s=-1$ to an NS5-brane and a $\left(1,k\right)$5-brane, respectively. We summarize the notation in table \ref{tab:BCNotation}. 

\begin{figure}
\begin{centering}
\includegraphics[scale=0.7]{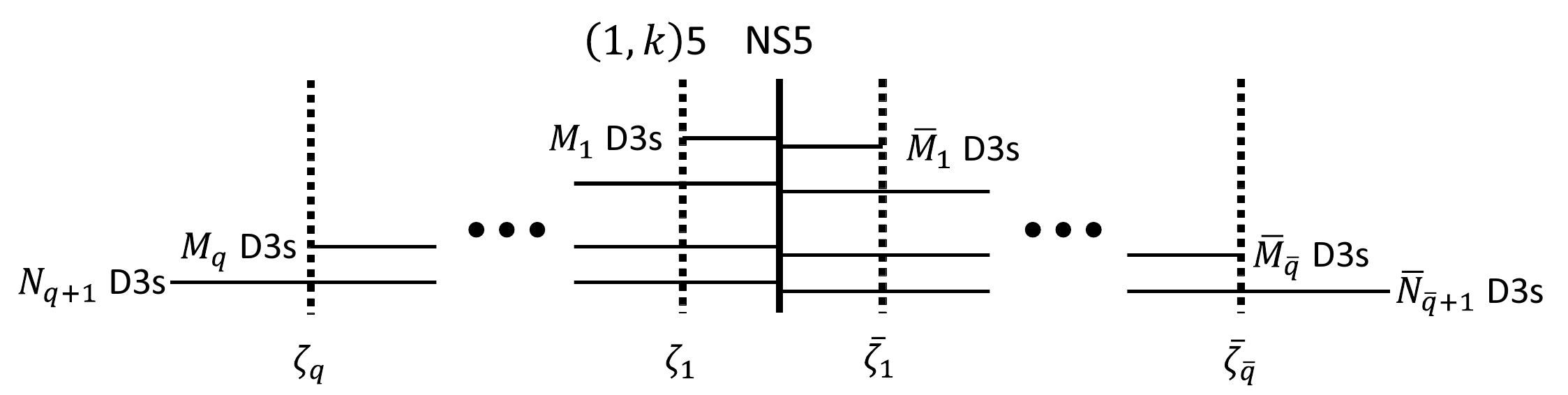}
\par\end{centering}
\caption{The local brane configuration shown in \eqref{eq:BCpart-Gen-dif}.\label{fig:BCpart-Gen}}
\end{figure}

In general, the brane configuration is the sequence of
\begin{equation}
\left\langle N_{q+1}\circ_{\zeta_{q}}N_{q}\cdots\circ_{\zeta_{1}}N_{1}\bullet_{\zeta}\bar{N}_{1}\circ_{\bar{\zeta}_{1}}\cdots\bar{N}_{\bar{q}}\circ_{\bar{\zeta}_{\bar{q}}}\bar{N}_{\bar{q}+1}\right\rangle .\label{eq:BCpart-Gen}
\end{equation}
This is the local brane configuration including just one NS5-brane. We assume that
\begin{equation}
N_{q+1}\leq N_{a}\quad\left(1\leq a\leq q\right),\quad\bar{N}_{\bar{q}+1}\leq\bar{N}_{a}\quad\left(1\leq a\leq\bar{q}\right),
\end{equation}
without loss of generality. For this brane configuration, we introduce the difference notation
\begin{equation}
\left\langle N_{q+1}\circ_{\zeta_{q},M_{q}}\cdots\circ_{\zeta_{1},M_{1}}\bullet_{\zeta}\circ_{\bar{\zeta}_{1},\bar{M}_{1}}\cdots\circ_{\bar{\zeta}_{\bar{q}},\bar{M}_{\bar{q}}}\bar{N}_{\bar{q}+1}\right\rangle ,\label{eq:BCpart-Gen-dif}
\end{equation}
where
\begin{equation}
M_{a}=N_{a}-N_{a+1},\quad\bar{M}_{\bar{a}}=\bar{N}_{\bar{a}}-\bar{N}_{\bar{a}+1}.\label{eq:DiffNot}
\end{equation}
The corresponding brane configuration is in figure \ref{fig:BCpart-Gen}.
We will see in section \ref{subsec:BadTh} that $M_{a}$ and $\bar{M}_{\bar{a}}$ are non-negative for restricting ourselves to ``good'' theories. Therefore, in this expression, $M$ on each $\left(1,k\right)$5-brane $\circ_{\zeta,M}$ denotes the number of the D3-branes which start with the $\left(1,k\right)$5-brane and end on the NS5-brane on the center.

\subsubsection{Restrictions for fractional D3-branes\label{subsec:BadTh}}

We will formally define a matrix model in the next section. However, in general, the integral of the matrix model is not always convergent. The condition whether the matrix model is divergent or not is equivalent to the condition whether the associated gauge theory is ``bad'' or not \cite{Kapustin:2010mh}. The good-ugly-bad classification was introduced in \cite{Gaiotto:2008ak}. To avoid the divergence, we do not consider ``bad'' theories. An ``ugly'' theory also has subtleties. There is a monopole operator whose scaling dimension is $\frac{1}{2}$ \cite{Gaiotto:2008ak}, which becomes a decoupled free sector in the IR fixed point. In this case, we have to multiply the dual matrix model by the corresponding factor so that a duality holds in the level of the matrix model \cite{Kapustin:2010mh}. In this paper, we also avoid to deal with ``ugly'' theories, and thus we focus on ``good'' theories.

The condition whether the worldvolume theory of a brane configuration is ``good'' or not depends on the linking numbers defined in \cite{Hanany:1996ie}. The linking number $\ell$ is defined for all 5-branes. In our setup, the linking number for a $\left(1,k\right)$5-brane is
\begin{equation}
\ell=n+kR,
\end{equation}
where $n$ denotes the number of the D3-branes ending on the $\left(1,k\right)$5-brane from the right minus the number of them ending from the left, and $R$ denotes the number of NS5-branes lying to the left. In order for the condition to be satisfied, we have to order the $\left(1,k\right)$5-branes so that the linking numbers are non-decreasing from left to right \cite{Gaiotto:2008ak}.\footnote{For a theory to be ``good'', the non-decreasing condition for the NS5-branes must also be imposed. However, our computation does not detect the divergence coming from this point. Therefore, in this paper we do not impose this condition.}

For our setup, there are two types of brane configurations we have to take care of. The non-decreasing constraint for these brane configurations leads to
\begin{align}
\left\langle N_{1}\circ N_{2}\circ N_{3}\right\rangle :\quad & N_{2}-N_{1}\leq N_{3}-N_{2},\nonumber \\
\left\langle N_{1}\circ N_{2}\bullet N_{3}\circ N_{4}\right\rangle :\quad & N_{2}-N_{1}\leq N_{4}-N_{3}+k.\label{eq:Good-Cond}
\end{align}
The restriction \eqref{eq:Good-Cond} for \eqref{eq:BCpart-Gen} is as follows. In the difference notation \eqref{eq:DiffNot}, the number of D3-branes in \eqref{eq:BCpart-Gen} should satisfy
\begin{align}
M_{a}\geq M_{a+1},\quad\bar{M}_{\bar{a}}\geq\bar{M}_{\bar{a}+1},\quad M_{1}+\bar{M}_{1}\leq k.\label{eq:BCCond-Good}
\end{align}
Since we cut the brane configuration where the number of D3-branes are (locally) minimum, $M_{a}$ and $\bar{M}_{\bar{a}}$ are non-negative. When $q=0$ or $\bar{q}=0$, we also impose
\begin{equation}
\bar{M}_{\bar{a}}\leq k,\quad M_{a}\leq k.\label{eq:BCCond-SB}
\end{equation}
When this restriction is not satisfied, the phenomena called duality cascade occurs \cite{Honda:2020uou}. The duality cascades relate the brane configuration which does and does not satisfy \eqref{eq:BCCond-SB}, otherwise the supersymmetry breaking occurs and the matrix model becomes zero.

\subsection{Matrix model\label{subsec:MM}}

After taking the low-energy limit, the worldvolume theory of the D3-branes is the three-dimensional ${\cal N}=4$ superconformal Chern-Simons theory with circular or linear quiver diagram. For each D3-brane segment between two 5-branes, a ${\rm U}\left(N\right)$ ${\cal N}=4$ vector multiplet appears, where $N$ denotes the number of D3-branes at the segment. The kinematic term of the vector multiplet depends on the 5-branes on both sides. For $\left\langle \circ N\bullet\right\rangle $ and $\left\langle \bullet N\circ\right\rangle $, the kinematic term is the Chern-Simons term with the Chern-Simons level $k$ and $-k$, respectively, while for $\left\langle \circ N\circ\right\rangle $ and $\left\langle \bullet N\bullet\right\rangle $, the vector multiplet becomes auxiliary \cite{Imamura:2008dt}. The relative displacement of the two 5-branes in the transverse direction gives the FI term. In addition, for each 5-brane, a bi-fundamental ${\cal N}=4$ hypermultiplet appears. Therefore, the worldvolume theory is the circular or the linear quiver gauge theory when there are regular D3-branes or not, respectively. In this paper, we call all of the worldvolume theories of brane configurations consisting of $p$ NS5-branes and $q$ $\left(1,k\right)$5-branes the ($p$,$q$) model.\footnote{The duality generated by the Hanany-Witten move holds in the level of the matrix model \cite{Assel:2014awa}. The Hanany-Witten move changes the order of an NS5-brane and a $\left(1,k\right)$5-brane. In this sense, we do not have to study the brane configurations which consist of the same number of 5-branes separately. The Hanany-Witten move is reviewed in appendix \ref{sec:HW}.}

The partition function on round three-sphere can be reduced to the integral over the Cartan subalgebra by using the supersymmetric localization technique \cite{Kapustin:2009kz,Hama:2010av}. We call the integral matrix model. The integrand consists of a classical factor and a 1-loop determinant factor. The Chern-Simons terms and the FI terms give rise to a classical contribution, while the vector multiplets and the hypermultiplets contribute to the 1-loop determinant factor. The classical contribution of the $\mathrm{U}\left(N\right)_{k}$ Chern-Simons term (here the subscript $k$ denotes the Chern-Simons level) and the FI term with the FI parameter $\zeta$ are
\begin{equation}
e^{i\pi k\sum_{n}^{N}\alpha_{n}^{2}},\quad e^{2\pi i\zeta\sum_{n}^{N}\alpha_{n}},
\end{equation}
where $\alpha_{n}$ are the corresponding eigenvalues. The $\mathrm{U}\left(N\right)$ vector multiplet contributes
\begin{equation}
\prod_{n<n'}^{N}\left(2\sinh\pi\left(\alpha_{n}-\alpha_{n'}\right)\right)^{2},
\end{equation}
while the $\mathrm{U}\left(N_{1}\right)\times\mathrm{U}\left(N_{2}\right)$ bi-fundamental hypermultiplet contributes
\begin{equation}
\frac{1}{\prod_{m}^{N_{1}}\prod_{n}^{N_{2}}2\cosh\pi\left(\alpha_{m}-\beta_{n}\right)}.
\end{equation}
The matrix model is the integral over the entire Cartan subalgebra divided by the order of the Weyl group. Namely, the integral is over the real line. The order of the Weyl group of ${\rm U}\left(N\right)$ is $N!$.

We saw that there is correspondence between the brane configurations and the matrix models. Furthermore, there is natural local correspondence between them. To see this, we explain a dictionary for obtaining the matrix model directly from the brane configuration. It needs two steps. First, we assign a factor to each 5-brane as follows:
\begin{align}
 & \left\langle N_{1}\bullet_{\zeta}N_{2}\right\rangle \rightarrow Z_{\zeta,\left(N_{1},N_{2}\right)}^{\left(+1\right)}\left(\alpha,\beta\right),\quad\left\langle N_{1}\circ_{\zeta}N_{2}\right\rangle \rightarrow Z_{\zeta,\left(N_{1},N_{2}\right)}^{\left(-1\right)}\left(\alpha,\beta\right),\label{eq:BCtoMM-Rule1}
\end{align}
where
\begin{align}
Z_{\zeta,\left(N_{1},N_{2}\right)}^{\left(+1\right)}\left(\alpha,\beta\right) & =e^{-2\pi i\zeta\left(\sum_{n}^{N_{1}}\alpha_{n}-\sum_{n}^{N_{2}}\beta_{n}\right)}Z_{N_{1},N_{2}}^{\lp}\left(\alpha,\beta\right),\nonumber \\
Z_{\zeta,\left(N_{1},N_{2}\right)}^{\left(-1\right)}\left(\alpha,\beta\right) & =i^{\frac{1}{2}\left(N_{1}^{2}-N_{2}^{2}\right)}e^{-i\pi k\left(\sum_{n}^{N_{1}}\alpha_{n}^{2}-\sum_{n}^{N_{2}}\beta_{n}^{2}\right)}e^{-2\pi i\zeta\left(\sum_{n}^{N_{1}}\alpha_{n}-\sum_{n}^{N_{2}}\beta_{n}\right)}Z_{N_{1},N_{2}}^{\lp}\left(\alpha,\beta\right),\label{eq:5factor-Def}
\end{align}
and
\begin{align}
Z_{N_{1},N_{2}}^{\lp}\left(\alpha,\beta\right) & =\frac{\prod_{m<m'}^{N_{1}}2\sinh\pi\left(\alpha_{m}-\alpha_{m'}\right)\prod_{n<n'}^{N_{2}}2\sinh\pi\left(\beta_{n}-\beta_{n'}\right)}{\prod_{m}^{N_{1}}\prod_{n}^{N_{2}}2\cosh\pi\left(\alpha_{m}-\beta_{n}\right)}.\label{eq:OneLoop}
\end{align}
Here, $\prod_{m<n}^{N}$ denotes $\prod_{m=1}^{N-1}\prod_{n=m+1}^{N}$. We call $Z_{\zeta,\left(N_{1},N_{2}\right)}^{\left(+1\right)}$ and $Z_{\zeta,\left(N_{1},N_{2}\right)}^{\left(-1\right)}$ the NS5-brane factor and the $\left(1,k\right)$5-brane factor, respectively. Note that the phase $i^{\frac{1}{2}\left(N_{1}^{2}-N_{2}^{2}\right)}$ is the natural generalization of the pure Chern-Simons theory \cite{Marino:2011nm} and the ABJ theory \cite{Drukker:2010nc}. Second, we glue all of the 5-brane factors with integrals and divide the integral by the order of the Weyl group. In this way, we can define the local matrix model corresponding to arbitrary local brane configuration. The local matrix model corresponding to \eqref{eq:BCpart-Gen} is
\begin{align}
Z_{\bm{\zeta},\boldsymbol{N}}^{\left(-q,+1,-\bar{q}\right)}\left(\alpha,\beta\right)= & \int\left(\prod_{a}^{q}\frac{d^{N_{a}}\lambda^{\left(a\right)}}{N_{a}!}\right)\left(\prod_{a}^{\bar{q}}\frac{d^{\bar{N}_{a}}\bar{\lambda}^{\left(a\right)}}{\bar{N}_{a}!}\right)\prod_{a}^{q}Z_{\zeta_{a}\left(N_{a+1},N_{a}\right)}^{\left(-1\right)}\left(\lambda^{\left(a+1\right)},\lambda^{\left(a\right)}\right)\nonumber \\
 & \times Z_{\zeta,\left(N_{1},\bar{N}_{1}\right)}^{\left(+1\right)}\left(\lambda^{\left(1\right)},\bar{\lambda}^{\left(1\right)}\right)\prod_{a}^{\bar{q}}Z_{\bar{\zeta}_{a},\left(\bar{N}_{a},\bar{N}_{a+1}\right)}^{\left(-1\right)}\left(\bar{\lambda}^{\left(a\right)},\bar{\lambda}^{\left(a+1\right)}\right),\label{eq:MM-Gen}
\end{align}
where
\begin{align}
\lambda^{\left(q+1\right)} & =\alpha,\quad\bar{\lambda}^{\left(\bar{q}+1\right)}=\beta,\nonumber \\
\bm{\zeta} & =\left(\zeta_{q},\ldots,\zeta_{1},\zeta,\bar{\zeta}_{1},\ldots,\bar{\zeta}_{\bar{q}}\right),\nonumber \\
\boldsymbol{N} & =\left(N_{q+1},\ldots,N_{1},\bar{N}_{1},\ldots,\bar{N}_{\bar{q}+1}\right).
\end{align}
The whole matrix model can be obtained by gluing the local matrix models (as the whole brane configuration can be obtained by gluing the local brane configurations).

We can flip all of the signs of the Chern-Simons levels $\pm k$ in the matrix model $Z$ simultaneously by taking the complex conjugate and flipping all of the signs of the FI parameters
\begin{equation}
\left.Z\right|_{k\rightarrow-k}=\left.Z^{*}\right|_{\bm{\zeta}\rightarrow-\bm{\zeta}}.\label{eq:MM-Conj}
\end{equation}
This corresponds to the fact that NS5-branes and $\left(1,k\right)$5-branes can be exchanged by using ${\rm SL}\left(2,\mathbb{Z}\right)$ transformation and the parity transformation.

\subsection{Quantum curve\label{subsec:QC}}

In this section, we review the relation between the matrix model and the quantum curve. We especially focus on the matrix models of the (2,2) model and the (2,4) model. We also review symmetry of the quantum curve and implications of the symmetry to the matrix model. 

The quantum curve is a operator which is a Laurent series of $\hat{Q}=e^{\hat{q}}$ and $\hat{P}=e^{\hat{p}}$. The notation of the quantum mechanics is as follows. The commutation relation of the position operator $\hat{q}$ and the momentum operator $\hat{p}$ is $\left[\hat{q},\hat{p}\right]=i\hbar$. In this paper, the reduced Planck constant is related to the Chern-Simons level as $\hbar=2\pi k$. $\ket{\cdot}$ denotes a position eigenvector. We also introduce a symbol $\kket{\cdot}$ which denotes a momentum eigenvector. The inner products of these vectors are
\begin{align}
 & \braket{q_{1}|q_{2}}=\delta\left(q_{1}-q_{2}\right),\quad\bbrakket{p_{1}|p_{2}}=\delta\left(p_{1}-p_{2}\right),\nonumber \\
 & \brakket{q|p}=\frac{1}{\sqrt{2\pi\hbar}}e^{\frac{i}{\hbar}pq},\quad\bbraket{p|q}=\frac{1}{\sqrt{2\pi\hbar}}e^{-\frac{i}{\hbar}pq}.\label{eq:Normalization}
\end{align}
The quantum curve is parameterized by the coefficients $\bm{e}$ as
\begin{equation}
\hat{H}_{\bm{e}}=\sum_{m,n\in\mathbb{Z}}e_{mn}\hat{Q}^{m}\hat{P}^{n}.
\end{equation}

There is an interesting relation between the matrix model associated to the $\mathcal{N}=4$ superconformal Chern-Simons theory and the quantum curve. All the known results strongly suggest that the matrix model can be written as
\begin{equation}
Z_{\bm{N}}^{\left\{ \bm{s}\right\} }=Z_{\bm{N}_{{\rm rel}}}^{\left\{ \bm{s}\right\} }\int\frac{d^{N_{{\rm min}}}x}{N_{{\rm min}}!}\det\left(\left[\braket{x_{m}|\hat{H}_{\bm{e}}^{-1}|x_{n}}\right]_{m,n}^{N_{{\rm min}}\times N_{{\rm min}}}\right),\label{eq:MM-QC}
\end{equation}
with a relation between $\bm{N}_{{\rm rel}}$ and $\bm{e}$. Here, $\bm{N}_{{\rm rel}}$ denotes the relative rank to the lowest rank $N_{{\rm min}}$, namely
\begin{equation}
N_{{\rm min}}=\min\bm{N},\quad\bm{N}_{{\rm rel}}=\bm{N}-N_{{\rm min}}\bm{1}.
\end{equation}
$\left[a_{mn}\right]_{m,n}^{M\times N}$ denotes an $M\times N$ matrix whose $\left(m,n\right)$ element is $a_{mn}$. It is remarkable that, in the right hand side, the $N_{{\rm min}}$ dependence appears only through the size of the matrix, and the prefactor $Z_{\bm{N}_{{\rm rel}}}^{\left\{ \bm{s}\right\} }$ and the quantum curve only depend on the relative ranks. The factor
\begin{equation}
Z_{N,\bm{e}}^{\left({\rm Fermi}\right)}=\int\frac{d^{N}x}{N!}\det\left(\left[\braket{x_{m}|\hat{H}_{\bm{e}}^{-1}|x_{n}}\right]_{m,n}^{N\times N}\right),\label{eq:FGpart}
\end{equation}
appearing in the right hand side of \eqref{eq:MM-QC} can be regarded as the partition function of the ideal Fermi gas with one-particle density matrix $\hat{H}_{\bm{e}}^{-1}$.

The type of relation \eqref{eq:MM-QC} was first discovered by using the Fermi gas formalism \cite{Marino:2011eh}. They found that the matrix models with nodes of equal ranks can be written in the form of \eqref{eq:MM-QC} (see also \cite{Moriyama:2014gxa}). After that, the Fermi gas formalism was applied for the ABJ theory, which is the ABJM theory with rank deformation \cite{Awata:2012jb,Honda:2013pea,Honda:2014npa,Kashaev:2015wia,Hatsuda:2016uqa}, and for more general rank deformed theories \cite{Kubo:2019ejc,Kubo:2020qed}.

Another procedure of identifying the relation between $Z_{\bm{N}}^{\left\{ \bm{s}\right\} }$ and $\hat{H}_{\bm{e}}$ was proposed in \cite{Kubo:2019ejc} for the (2,2) model. This procedure was also extended for the (2,4) model in \cite{Furukawa:2020cjp}. These procedure directly relate $Z_{\bm{N}}^{\left\{ \bm{s}\right\} }$ and $\hat{H}_{\bm{e}}$ by assuming \eqref{eq:MM-QC}. In this procedure, we cannot proof the parameter identification between $\bm{N}_{{\rm rel}}$ and $\bm{e}$ due to lack of the explicit computation such as Fermi gas approach.

\subsubsection{Brane transitions from quantum curve\label{subsec:Implications}}

The symmetries of the quantum curves corresponding to the (2,2) model and the (2,4) model were studied in \cite{Kubo:2018cqw}. The symmetry is defined such that some quantum curves which are the same up to similarity transformations are the dual each other. Since the Fermi gas factor $Z_{N,\bm{e}}^{\left({\rm Fermi}\right)}$ in \eqref{eq:FGpart} can be written in a combination of the trace of $\hat{H}_{\bm{e}}^{-n}$, this factor is invariant under the similarity transformations. Therefore, the symmetry of the quantum curve immediately provides the equality among the Fermi gas factors if the relation between $\bm{N}_{{\rm rel}}$ and $\bm{e}$ is known. On the other hand, the quantum curve does not have any information about the prefactor $Z_{\bm{N}_{{\rm rel}}}^{\left\{ \bm{s}\right\} }$. We will ignore this point for now and focusing on the fact that the symmetry of the quantum curve implies the dualities of gauge theories since the equality of the partition functions implies the duality. To interpret the symmetry to the dualities, we have to know the relation between $\bm{N}_{{\rm rel}}$, which is the parameter of the gauge theory, and $\bm{e}$, which is the coefficients of the quantum curve. If we know the relation, since the symmetry of the quantum curve relates different values of $\bm{e}$, we obtain the set of gauge theories and brane configurations corresponding to $\bm{N}_{{\rm rel}}$. Note that, in the duality, the lowest rank $N_{{\rm min}}$ is kept invariant. We also keep the type and the number of 5-branes. In summary, the symmetry of the quantum curve implies the relations in the gauge theories and brane configurations where the lowest rank and the number of 5-branes are fixed.

The parameter identification was performed for the (2,2) model in \cite{Kubo:2019ejc} and for the (2,4) model in \cite{Furukawa:2020cjp} when all of the FI parameters are zero. They found that some symmetries of the quantum curves correspond to the Hanany-Witten transition, while there exist symmetries which can not be generated by the Hanany-Witten transition. For the (2,2) model,
\begin{equation}
\left\langle N_{1}\circ N_{2}\bullet N_{3}\circ N_{4}\bullet\right\rangle ^{\mathrm{P}}\sim\left\langle N_{1}\circ N_{3}\bullet N_{2}\circ N_{4}\bullet\right\rangle ^{\mathrm{P}},\label{eq:ExcepSymD5}
\end{equation}
is the typical symmetry. Here $\sim$ denotes that the relation comes from the symmetry of the quantum curve. For the (2,4) model,
\begin{align}
\left\langle N_{1}\bullet N_{2}\circ N_{3}\bullet N_{4}\circ N_{5}\circ N_{6}\circ\right\rangle ^{\mathrm{P}} & \sim\left\langle N_{1}\bullet N_{3}\circ N_{2}\bullet N_{4}\circ N_{5}\circ N_{6}\circ\right\rangle ^{\mathrm{P}},\nonumber \\
\left\langle N_{1}\bullet N_{2}\circ N_{3}\circ N_{4}\bullet N_{5}\circ N_{6}\circ\right\rangle ^{\mathrm{P}} & \sim\left\langle N_{2}\bullet N_{1}\circ N_{3}\circ N_{5}\bullet N_{4}\circ N_{6}\circ\right\rangle ^{\mathrm{P}},\label{eq:ExcepSymE7}
\end{align}
are the typical symmetries when 
\begin{align}
N_{1}-N_{2} & =N_{3}-N_{4}+k,\nonumber \\
N_{1}-N_{2} & =N_{4}-N_{5},\label{eq:Balance-Cond}
\end{align}
respectively. These conditions are called the balanced condition \cite{Furukawa:2020cjp}. In \cite{Furukawa:2020cjp}, motivated by \eqref{eq:ExcepSymD5} and \eqref{eq:ExcepSymE7} (with some additional cases), a local brane transition
\begin{equation}
\left\langle N_{1}\bullet N_{2}\circ N_{3}\bullet N_{4}\right\rangle \sim\left\langle N_{1}\bullet N_{3}\circ N_{2}\bullet N_{4}\right\rangle ,\label{eq:FMN-trans}
\end{equation}
was also conjectured. In section \ref{sec:Res-QC}, we will see that the relations we find include \eqref{eq:ExcepSymD5} and \eqref{eq:ExcepSymE7} and overlap with \eqref{eq:FMN-trans}.

\section{Two NS5-branes\label{sec:2q-model}}

In the previous section, we explained the relation between the brane configuration, the matrix model and the quantum curve. Especially, in section \ref{subsec:QC}, we saw that the symmetry of the quantum curve implies non-trivial relations among the matrix models or brane configurations. We explained that there are relations which cannot be explained in the Hanany-Witten transition as shown in \eqref{eq:ExcepSymD5} and \eqref{eq:ExcepSymE7}. The above story, however, has some problems as explained in the introduction. Here, we again explain these points by taking into account what we explained in the previous section. First, the implications for the brane transitions are restricted to the (2,2) model and the (2,4) model. Second, though the 5-branes have the relative FI parameters as the shift to the transverse space, the quantum curves do not have information about them. Third, though the procedure of the identification employed for the (2,2) model and the (2,4) model provides the conjectured relation of the parameters, there is no proof yet. Fourth, the symmetries of the quantum curve say nothing about the prefactor $Z_{\bm{N}_{{\rm rel}}}^{\left\{ \bm{s}\right\} }$ in \eqref{eq:MM-QC}.

In this paper, we resolve these problems by directly studying the matrix model. In this section, we focus on the case when there are two NS5-branes (and one NS5-brane as a special case). More general cases will be studied in the next section. In section \ref{subsec:Duality-BC}, we present new relations as a brane transition. In section \ref{subsec:Duality-MM}, we show exact relations among the matrix models and a physical interpretation of them. This is our main result. In section \ref{subsec:Proof}, we prove the relations by employing the improved Fermi gas formalism.

\subsection{Dualities as brane transitions\label{subsec:Duality-BC}}

In this section, we present new dualities in the brane picture. As we will show in section \ref{subsec:Duality-MM}, the dualities hold only after adding the decoupled sector. However, we postpone the study of this point to the next section, and instead we show the dualities as brane transitions in this section. One reason why we first show the dualities in the brane picture is that the dualities are quite simple in the brane picture.

We start with the brane configuration consisting of two NS5-branes and the arbitrary number of $\left(1,k\right)$5-branes:
\begin{equation}
\left\langle N\circ_{\zeta_{q}}N_{q}\cdots\circ_{\zeta_{1}}N_{1}\bullet_{\zeta}\bar{N}_{1}\circ_{\bar{\zeta}_{1}}\cdots\bar{N}_{\bar{q}}\circ_{\bar{\zeta}_{\bar{q}}}N'\circ_{\zeta_{q'}'}N_{q'}'\cdots\circ_{\zeta_{1}'}N_{1}'\bullet_{\zeta'}\bar{N}_{1}'\circ_{\bar{\zeta}_{1}'}\cdots\bar{N}_{\bar{q}'}'\circ_{\bar{\zeta}_{\bar{q}'}'}\right\rangle ^{\mathrm{P}}.\label{eq:BC-Gen}
\end{equation}
This is the brane picture of (2,$q+\bar{q}+q'+\bar{q}'$) model.

To explain brane transitions we found, it is convenient to move on the difference notation \eqref{eq:DiffNot}. There are two parts where all 5-branes are $\left(1,k\right)$5-branes, and for each part, there is a point where the number of the D3-branes are minimum. Here we assume that $N$ and $N'$ are the minimum, namely
\begin{equation}
N\leq N_{a},\quad N\leq\bar{N}'_{\bar{a}'},\quad N'\leq N_{a'}',\quad N'\leq\bar{N}_{\bar{a}},
\end{equation}
We cut the brane configuration at the two points, and then we obtain two local brane configurations
\begin{align}
 & \left\langle N_{q+1}\circ_{\zeta_{q}}N_{q}\cdots\circ_{\zeta_{1}}N_{1}\bullet_{\zeta}\bar{N}_{1}\circ_{\bar{\zeta}_{1}}\cdots\bar{N}_{\bar{q}}\circ_{\bar{\zeta}_{\bar{q}}}\bar{N}_{\bar{q}+1}\right\rangle ,\nonumber \\
 & \left\langle N_{q'+1}'\circ_{\zeta_{q'}'}N_{q'}'\cdots\circ_{\zeta_{1}'}N_{1}'\bullet_{\zeta'}\bar{N}_{1}'\circ_{\bar{\zeta}_{1}'}\cdots\bar{N}_{\bar{q}'}'\circ_{\bar{\zeta}_{\bar{q}'}'}\bar{N}_{\bar{q}'+1}'\right\rangle ,
\end{align}
where $N_{q+1}=\bar{N}_{\bar{q}'+1}'=N$ and $\bar{N}_{\bar{q}+1}=N_{q'+1}'=N'$. For each local brane configuration, we can write it in the difference notation \eqref{eq:BCpart-Gen-dif}. Therefore, in the difference notation, \eqref{eq:BC-Gen} can be written as
\begin{equation}
\left\langle N\circ_{\zeta_{q},M_{q}}\cdots\circ_{\zeta_{1},M_{1}}\bullet_{\zeta}\circ_{\bar{\zeta}_{1},\bar{M}_{1}}\cdots\circ_{\bar{\zeta}_{\bar{q}},\bar{M}_{\bar{q}}}N'\circ_{\zeta_{q'}',M_{q'}'}\cdots\circ_{\zeta_{1}',M_{1}'}\bullet_{\zeta'}\circ_{\bar{\zeta}'_{1},\bar{M}_{1}'}\cdots\circ_{\bar{\zeta}'_{\bar{q}'},\bar{M}'_{\bar{q}'}}\right\rangle ^{\mathrm{P}}.\label{eq:BC-Gen-Diff}
\end{equation}
In this notation, we find the brane transition
\begin{align}
 & \left\langle N\circ_{\zeta_{q},M_{q}}\cdots\circ_{\zeta_{1},M_{1}}\bullet_{\zeta}\circ_{\bar{\zeta}_{1},\bar{M}_{1}}\cdots\circ_{\bar{\zeta}_{\bar{q}},\bar{M}_{\bar{q}}}N'\circ_{\zeta_{q'}',M_{q'}'}\cdots\circ_{\zeta_{1}',M_{1}'}\bullet_{\zeta'}\circ_{\bar{\zeta}_{1}',\bar{M}_{1}'}\cdots\circ_{\bar{\zeta}_{\bar{q}'}',\bar{M}_{\bar{q}'}'}\right\rangle ^{\mathrm{P}}\nonumber \\
 & \leftrightarrow\left\langle N\circ_{\eta_{r},L_{r}}\cdots\circ_{\eta_{1},L_{1}}\bullet_{\zeta}\circ_{\bar{\eta}_{1},\bar{L}_{1}}\cdots\circ_{\bar{\eta}_{\bar{r}},\bar{L}_{\bar{r}}}N'\circ_{\eta_{r'}',L_{r'}'}\cdots\circ_{\eta_{1}',L_{1}'}\bullet_{\zeta'}\circ_{\bar{\eta}_{1}',\bar{L}_{1}'}\cdots\circ_{\bar{\eta}_{\bar{r}'}',\bar{L}_{\bar{r}'}'}\right\rangle ^{\mathrm{P}},\label{eq:Duality-BC}
\end{align}
where the set of $\left(1,k\right)$5-branes in each region is the same:
\begin{align}
\left\{ \circ_{\zeta_{1},M_{1}},\cdots,\circ_{\zeta_{q},M_{q}},\circ_{\bar{\zeta}_{1}',\bar{M}_{1}'},\cdots,\circ_{\bar{\zeta}_{\bar{q}'}',\bar{M}_{\bar{q}'}'}\right\}  & =\left\{ \circ_{\eta_{1},L_{1}},\cdots,\circ_{\eta_{r},L_{r}},\circ_{\bar{\eta}_{1}',\bar{L}_{1}'},\cdots,\circ_{\bar{\eta}_{\bar{r}'}',\bar{L}_{\bar{r}'}'}\right\} ,\nonumber \\
\left\{ \circ_{\bar{\zeta}_{1},\bar{M}_{1}},\cdots,\circ_{\bar{\zeta}_{\bar{q}},\bar{M}_{\bar{q}}},\circ_{\zeta_{1}',M_{1}'},\cdots,\circ_{\zeta_{q'}',M_{q'}'}\right\}  & =\left\{ \circ_{\bar{\eta}_{1},\bar{L}_{1}},\cdots,\circ_{\bar{\eta}_{\bar{r}},\bar{L}_{\bar{r}}},\circ_{\eta_{1}',L_{1}'},\cdots,\circ_{\eta_{r'}',L_{r'}'}\right\} ,\label{eq:Duality-BC-Set}
\end{align}
(so that $q+\bar{q}'=r+\bar{r}'$ and $\bar{q}+q'=\bar{r}+r'$). Namely, we find that the fractional D3-branes can change the NS5-brane where they end.\footnote{This does not mean that the orientation is changed. They are both the ordinary D3-branes.} 
\begin{figure}
\begin{centering}
\includegraphics[scale=0.7]{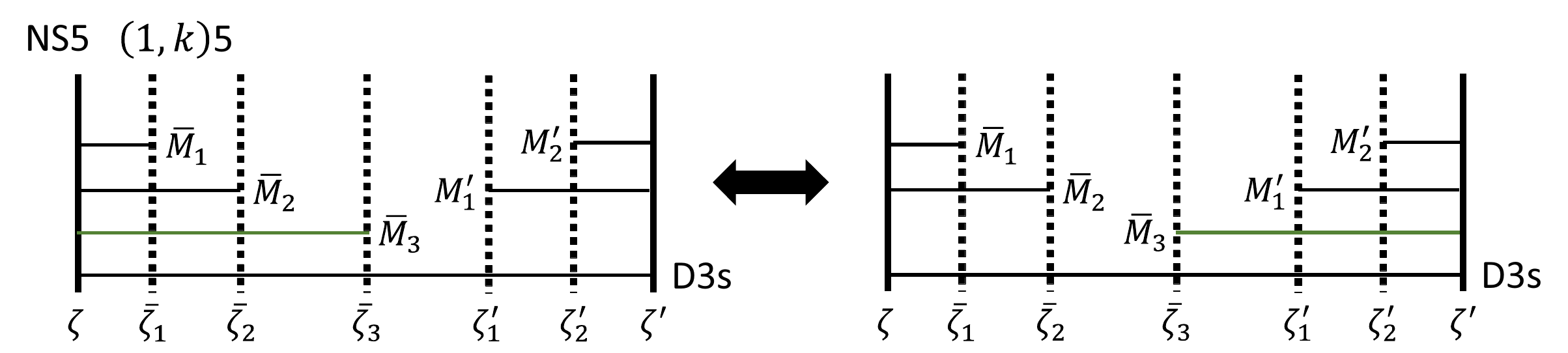}
\par\end{centering}
\caption{An example of the brane transition in \eqref{eq:Duality-BC} when $\bar{q}=3$ and $q'=2$. We only write the center of the brane configuration between the two NS5-branes. The fractional D3-branes represented by the horizontal green line change the NS5-brane where they end before and after the transition. The FI parameters and the total number of the fractional D3-branes are not changed.\label{fig:BC-trans}}
\end{figure}
 See figure \ref{fig:BC-trans} as an example of this transition. In contrast to the gauge theory picture, the transition in the brane picture is quite simple. This is one motivation to interpret the dualities as the brane transitions. 

As a simple consistency check of the dualities, we emphasize that the moduli spaces of dual theories which describe the motion of the M2-branes match. After T-dualizing and uplifting the type IIB brane configuration of the ($p$,$q$) model to the M-theory, the regular D3-branes become the M2-branes probing $\left(\mathbb{C}^{2}/\mathbb{Z}_{p}\times\mathbb{C}^{2}/\mathbb{Z}_{q}\right)/\mathbb{Z}_{k}$ \cite{Imamura:2008nn}. This geometry depends only on $k$, $p$, $q$ (and the lowest rank, which is the number of freely moving M2-branes). These values of the dual theories are the same.

Note that once the NS5-brane on which the D3-branes end is chosen, as explained in section \ref{subsec:BadTh}, we have to order the $\left(1,k\right)$5-branes so that the ``good'' condition \eqref{eq:BCCond-Good} is satisfied. Sometimes \eqref{eq:BCCond-Good} is never satisfied even if we start with the brane configuration which satisfies the condition. Therefore, the brane transition \eqref{eq:Duality-BC} is only for the case when both of the sides satisfy the ``good'' condition. On the other hand, the condition from the duality cascade \eqref{eq:BCCond-SB} is automatically satisfied when it is satisfied for the original brane configurations.

In general, if both of the dual theories are the worldvolume theories of brane configurations, it is expected that there are corresponding dualities or symmetries in the string theory. The typical symmetries in the Hanany-Witten setup are the Hanany-Witten transition and the ${\rm SL}\left(2,\mathbb{Z}\right)$ transformation. However, it is easy to see that the Hanany-Witten transition does not generate our brane transitions. Furthermore, in our setup, the ${\rm SL}\left(2,\mathbb{Z}\right)$ transformation only exchanges the NS5-branes and the $\left(1,k\right)$5-branes (if we also perform the parity transformation $k\rightarrow-k$). In this sense, at least when we focus on the Hanany-Witten setup, our brane transition is new.

Finally, let us consider a simpler case. We set $\bar{q}=q'=0$ and identify the two NS5-branes $\bullet_{\zeta}$ and $\bullet_{\zeta'}$ in \eqref{eq:BC-Gen-Diff}. Namely, we consider the brane configuration
\begin{equation}
\left\langle N\circ_{\zeta_{q},M_{q}}\cdots\circ_{\zeta_{1},M_{1}}\bullet_{\zeta}\circ_{\bar{\zeta}'_{1},\bar{M}_{1}'}\cdots\circ_{\bar{\zeta}'_{\bar{q}'},\bar{M}'_{\bar{q}'}}\right\rangle ^{\mathrm{P}}.\label{eq:BC-NS1-Diff}
\end{equation}
The duality for this brane configuration can be read off from \eqref{eq:Duality-BC} and \eqref{eq:Duality-BC-Set}:
\begin{align}
 & \left\langle N\circ_{\zeta_{q},M_{q}}\cdots\circ_{\zeta_{1},M_{1}}\bullet_{\zeta}\circ_{\bar{\zeta}_{1}',\bar{M}_{1}'}\cdots\circ_{\bar{\zeta}_{\bar{q}'}',\bar{M}_{\bar{q}'}'}\right\rangle ^{\mathrm{P}}\leftrightarrow\left\langle N\circ_{\eta_{r},L_{r}}\cdots\circ_{\eta_{1},L_{1}}\bullet_{\zeta}\circ_{\bar{\eta}_{1}',\bar{L}_{1}'}\cdots\circ_{\bar{\eta}_{\bar{r}'}',\bar{L}_{\bar{r}'}'}\right\rangle ^{\mathrm{P}},\label{eq:Duality-NS1-BC}
\end{align}
where the set of the $\left(1,k\right)$5-branes is the same:
\begin{align}
\left\{ \circ_{\zeta_{1},M_{1}},\cdots,\circ_{\zeta_{q},M_{q}},\circ_{\bar{\zeta}_{1}',\bar{M}_{1}'},\cdots,\circ_{\bar{\zeta}_{\bar{q}'}',\bar{M}_{\bar{q}'}'}\right\}  & =\left\{ \circ_{\eta_{1},L_{1}},\cdots,\circ_{\eta_{r},L_{r}},\circ_{\bar{\eta}_{1}',\bar{L}_{1}'},\cdots,\circ_{\bar{\eta}_{\bar{r}'}',\bar{L}_{\bar{r}'}'}\right\} .
\end{align}
The duality \eqref{eq:21Dual-Ex} is an example of this duality. We will see this in the next section.

\subsection{Dualities in quiver gauge theories\label{subsec:Duality-MM}}

In this section, we study the dualities presented in the previous section in more detail by using the matrix models. The matrix model corresponding to the whole brane configuration \eqref{eq:BC-Gen} is
\begin{equation}
Z_{\left(\boldsymbol{N},\bm{N}'\right)}=\int\frac{d^{N}\alpha}{N!}\frac{d^{N'}\beta}{N'!}Z_{\bm{\zeta},\boldsymbol{N}}^{\left(-q,+1,-\bar{q}\right)}\left(\alpha,\beta\right)Z_{\bm{\zeta}',\boldsymbol{N}'}^{\left(-q',+1,-\bar{q}'\right)}\left(\beta,\alpha\right).\label{eq:Zhole-Gen-Def}
\end{equation}
For \eqref{eq:Duality-BC}, in terms of the matrix model, we find the functional relation
\begin{align}
\frac{Z_{\left(\boldsymbol{N},\bm{N}'\right)}^{\left({\rm Left}\right)}}{\Omega_{\bm{\zeta},\bm{N}}^{q,\bar{q}}\Omega_{\bm{\zeta}',\bm{N}'}^{q',\bar{q}'}Z_{\left(\bm{\zeta}_{{\rm L}};\bm{\zeta}_{{\rm R}}\right),\left(\bm{M};\bar{\bm{M}}\right)}^{\left(\bifund\right)}Z_{\left(\bm{\zeta}_{{\rm L}}';\bm{\zeta}_{{\rm R}}'\right),\left(\bm{M}';\bar{\bm{M}}'\right)}^{\left(\bifund\right)}}=\frac{Z_{\left(\bm{K},\bm{K}'\right)}^{\left({\rm Right}\right)}}{\Omega_{\bm{\eta},\bm{K}}^{r,\bar{r}}\Omega_{\bm{\eta}',\bm{K}'}^{r',\bar{r}'}Z_{\left(\bm{\eta}_{{\rm L}};\bm{\eta}_{{\rm R}}\right),\left(\bm{L};\bar{\bm{L}}\right)}^{\left(\bifund\right)}Z_{\left(\bm{\eta}_{{\rm L}}';\bm{\eta}_{{\rm R}}'\right),\left(\bm{L}';\bar{\bm{L}}'\right)}^{\left(\bifund\right)}},\label{eq:Duality-MM}
\end{align}
where $Z_{\left(\boldsymbol{N},\bm{N}'\right)}^{\left({\rm Left}\right)}$ and $Z_{\left(\bm{K},\bm{K}'\right)}^{\left({\rm Right}\right)}$ are the matrix models corresponding to the left hand side and the right hand side of \eqref{eq:Duality-BC}, respectively. $\bm{K}$ and $\bm{K}'$ are the ranks of the right side brane configuration:
\begin{align}
\bm{K} & =\left(N,N+L_{q}\ldots,N+L_{1},N'+\bar{L}_{1},\ldots,N'+\bar{L}_{\bar{q}},N'\right),\nonumber \\
\bm{K}' & =\left(N',N'+L_{q}'\ldots,N'+L_{1}',N+\bar{L}_{1}',\ldots,N+\bar{L}_{\bar{q}}',N\right).
\end{align}
The factors in the denominator are defined in appendix \ref{sec:FGF_loc}. The proof of this relation is in section \ref{subsec:Proof}.

To consider the physical impression of this functional identity, we explain the physical interpretations of the factors appearing in the denominator. The phase factor $\Omega_{\bm{\zeta},\bm{N}}^{q,\bar{q}}$ is the framing factor, at least when all of the FI parameters are zero \cite{Witten:1988hf,Marino:2011nm}. The framing factor comes from the fact that the localization technique implicitly choose the non-trivial framing \cite{Kapustin:2009kz}.

The bi-fundamental factor $Z_{\left(\bm{\zeta}_{{\rm L}};\bm{\zeta}_{{\rm R}}\right),\left(\bm{M};\bar{\bm{M}}\right)}^{\left(\bifund\right)}$ is more physical. The physical interpretation of this factor is in appendix \ref{subsec:Factor-Meaning}. It captures the dynamics of fundamental strings stretched between the fractional D3-branes. In other words, the bi-fundamental factor only has the information about the fractional D3-branes. Notice that the Chern-Simons factor $Z_{\left(\bm{M},\bar{\bm{M}}\right)}^{\left(\cstot\right)}$ in \eqref{eq:FGF-Res} also captures the dynamics of fundamental strings on the fractional D3-branes as explained in appendix \ref{subsec:Factor-Meaning}, and thus it only has the information about the fractional D3-branes. Therefore, the Fermi gas formalism \eqref{eq:FGF-Res} shows that the functions associated to the fractional D3-branes factorize. This implies that the dynamics on the fractional D3-branes decouples in the IR fixed point.

This argument is natural from the viewpoint of the brane configuration. This is because the brane configuration consisting of only the fractional D3-branes can be obtained by separating the regular D3-branes. In the gauge theory, this corresponds to going to a generic point on the moduli space, where at the IR fixed point we get $N$ copies of a ${\rm U}\left(1\right)^{r}$ theory, together with a gauge theory corresponding to the fractional D3-branes \cite{Aharony:2008gk}.

Notice that, in the complete Fermi gas formalism \eqref{eq:MM-QC}, the claim that the dynamics purely coming form the fractional D3-branes decouples in the IR is more natural. This is because the prefactor independent of the lowest rank is the partition function of the worldvolume theory which is obtained by removing the regular D3-branes. We can easily see this by setting the lowest rank $N_{{\rm min}}$ to zero. Therefore, the complete Fermi gas form clearly shows that the factors corresponding to the dynamics of fractional D3-branes factorizes. Here, we do not claim that all of the dynamics related to the fractional D3-branes decouple. Indeed, the Fermi gas partition function also captures the fractional D3-branes.

This looks analogous to the ``ugly-good'' duality, where the monopole operator whose dimension is $\frac{1}{2}$ decouples in the IR fixed point and becomes free hypermultiplet \cite{Gaiotto:2008ak} (although the decoupled theory is an interacting SCFT in our case). In that case, we need to add the decoupled sector to the dual ``good'' theory. This tells us that we have to give the same prescription in this case. 

This can be explicitly seen in the level of the matrix models. We assume $N\leq N'$ for simplicity. By using \eqref{eq:Duality-MM} and the fact that the prefactors depend only on the relative ranks (see appendix \ref{sec:FGF_loc}), we obtain
\begin{equation}
Z_{\left(\boldsymbol{N},\bm{N}'\right)}^{\left({\rm Left}\right)}Z_{\left(\bm{K}-N\bm{1},\bm{K}'-N\bm{1}\right)}^{\left({\rm Right}\right)}=Z_{\left(\bm{K},\bm{K}'\right)}^{\left({\rm Right}\right)}Z_{\left(\bm{N}-N\bm{1},\bm{N}'-N\bm{1}\right)}^{\left({\rm Left}\right)}.\label{eq:CL-Dual-MM}
\end{equation}
Here, the second matrix models in both the sides are the matrix models associated to the fractional D3-branes since all of the ranks are substituted by the lowest rank $N$. The corresponding theories have linear quivers since there are no regular D3-branes.\footnote{If there are more than one block of fractional branes, the decoupled sector would become a product of decoupled linear quiver gauge theories. In this case, the matrix model for the decoupled sector $Z_{\left(\bm{N}-N\bm{1},\bm{N}'-N\bm{1}\right)}$ becomes the product of the corresponding matrix models for the linear quiver theories.} Therefore, this relation implies that there are dualities in the three-dimensional $\mathcal{N}=4$ superconformal Chern-Simons theories which are the product of a circular quiver gauge theory and a decoupled linear quiver gauge theory. In brane picture, the worldvolume theories of
\begin{align}
 & \left\langle N\circ_{\zeta_{q},M_{q}}\cdots\circ_{\zeta_{1},M_{1}}\bullet_{\zeta}\circ_{\bar{\zeta}_{1},\bar{M}_{1}}\cdots\circ_{\bar{\zeta}_{\bar{q}},\bar{M}_{\bar{q}}}N'\circ_{\zeta_{q'}',M_{q'}'}\cdots\circ_{\zeta_{1}',M_{1}'}\bullet_{\zeta'}\circ_{\bar{\zeta}_{1}',\bar{M}_{1}'}\cdots\circ_{\bar{\zeta}_{\bar{q}'}',\bar{M}_{\bar{q}'}'}\right\rangle ^{\mathrm{P}}\nonumber \\
 & \oplus\left\langle 0\circ_{\eta_{r},L_{r}}\cdots\circ_{\eta_{1},L_{1}}\bullet_{\zeta}\circ_{\bar{\eta}_{1},\bar{L}_{1}}\cdots\circ_{\bar{\eta}_{\bar{r}},\bar{L}_{\bar{r}}}N'-N\circ_{\eta_{r'}',L_{r'}'}\cdots\circ_{\eta_{1}',L_{1}'}\bullet_{\zeta'}\circ_{\bar{\eta}_{1}',\bar{L}_{1}'}\cdots\circ_{\bar{\eta}_{\bar{r}'}',\bar{L}_{\bar{r}'}'}0\right\rangle ,\label{eq:BCdualL}
\end{align}
and
\begin{align}
 & \left\langle N\circ_{\eta_{r},L_{r}}\cdots\circ_{\eta_{1},L_{1}}\bullet_{\zeta}\circ_{\bar{\eta}_{1},\bar{L}_{1}}\cdots\circ_{\bar{\eta}_{\bar{r}},\bar{L}_{\bar{r}}}N'\circ_{\eta_{r'}',L_{r'}'}\cdots\circ_{\eta_{1}',L_{1}'}\bullet_{\zeta'}\circ_{\bar{\eta}_{1}',\bar{L}_{1}'}\cdots\circ_{\bar{\eta}_{\bar{r}'}',\bar{L}_{\bar{r}'}'}\right\rangle ^{\mathrm{P}}\nonumber \\
 & \oplus\left\langle 0\circ_{\zeta_{q},M_{q}}\cdots\circ_{\zeta_{1},M_{1}}\bullet_{\zeta}\circ_{\bar{\zeta}_{1},\bar{M}_{1}}\cdots\circ_{\bar{\zeta}_{\bar{q}},\bar{M}_{\bar{q}}}N'-N\circ_{\zeta_{q'}',M_{q'}'}\cdots\circ_{\zeta_{1}',M_{1}'}\bullet_{\zeta'}\circ_{\bar{\zeta}_{1}',\bar{M}_{1}'}\cdots\circ_{\bar{\zeta}_{\bar{q}'}',\bar{M}_{\bar{q}'}'}0\right\rangle ,\label{eq:BCdualR}
\end{align}
are the dual each other. In this picture, it is easy to see that the additional theories are linear quiver gauge theories.

Notice that, based on the above argument and the conjectured Fermi gas formalism \eqref{eq:MM-QC}, the symmetries of the quantum curve immediately mean the same form relations among the corresponding matrix models. Namely, since the symmetries of the quantum curves mean the equalities of the Fermi gas factor $Z_{N,\bm{e}}^{\left({\rm Fermi}\right)}$ in \eqref{eq:FGpart} as explained in \ref{subsec:Implications}, we obtain
\begin{equation}
Z_{N\bm{1}+\bm{N}_{{\rm rel}}}^{\left\{ \bm{s}\right\} }Z_{\bm{N}_{{\rm rel}}'}^{\left\{ \bm{s}\right\} }=Z_{N\bm{1}+\bm{N}_{{\rm rel}}'}^{\left\{ \bm{s}\right\} }Z_{\bm{N}_{{\rm rel}}}^{\left\{ \bm{s}\right\} },
\end{equation}
where the quantum curves corresponding to $\bm{N}_{{\rm rel}}'$ and $\bm{N}_{{\rm rel}}'$ live in the same orbit of the symmetries. This argument further shows that there exist dualities corresponding to all the symmetries of the quantum curves if we assume \eqref{eq:MM-QC}. This is because the prefactor is the matrix model corresponding to the linear quiver gauge theory obtained after removing the regular D3-branes. Therefore, the symmetries of the quantum curves strongly imply the dualities with the decoupled linear quiver gauge theory.

We finally explain our results for two simple cases. First, we consider the (2,2) model. In the brane configuration, as a special case of \eqref{eq:Duality-BC}, we find
\begin{equation}
\left\langle N\circ_{\zeta_{1}}N+M_{1}\bullet_{\zeta}N'\circ_{\zeta_{1}'}N'+M_{1}'\bullet_{\zeta'}\right\rangle ^{\mathrm{P}}\leftrightarrow\left\langle N\circ_{\zeta_{1}}N+M_{1}\bullet_{\zeta}N'+M_{1}'\circ_{\zeta_{1}'}N'\bullet_{\zeta'}\right\rangle ^{\mathrm{P}}.\label{eq:Duality-BC-22-ex}
\end{equation}
or equivalently,
\begin{equation}
\left\langle N\circ_{\zeta_{1},M_{1}}\bullet_{\zeta}N'\circ_{\zeta_{1}',M_{1}'}\bullet_{\zeta'}\right\rangle ^{\mathrm{P}}\leftrightarrow\left\langle N\circ_{\zeta_{1},M_{1}}\bullet_{\zeta}\circ_{\zeta_{1}',M_{1}'}N'\bullet_{\zeta'}\right\rangle ^{\mathrm{P}}.
\end{equation}
This relation is clearly not generated by the Hanany-Witten transition. In the matrix model, as a special case of \eqref{eq:Duality-MM}, we have
\begin{equation}
\frac{Z_{\left(N,N+M_{1},N',N'+M_{1}'\right)}^{\left({\rm Left}\right)}}{\Omega_{\left(\zeta_{1},\zeta\right),\left(N,N+M_{1},N'\right)}^{1,0}\Omega_{\left(\zeta_{1}',\zeta'\right),\left(N',N'+M_{1}',N\right)}^{1,0}}=\frac{Z_{\left(N,N+M_{1},N'+M_{1}',N'\right)}^{\left({\rm Right}\right)}}{\Omega_{\left(\zeta_{1},\zeta,\zeta_{1}'\right),\left(N,N+M_{1},N'+M_{1}',N'\right)}^{1,1}\Omega_{\left(\zeta'\right),\left(N',N\right)}^{0,0}Z_{M_{1},M_{1}'}^{\left(\matter\right)}\left(\zeta_{1}-\zeta_{1}'\right)}.\label{eq:Duality-MM-22-ex}
\end{equation}
The discrepancy except for the phase is $Z_{M_{1},M_{1}'}^{\left(\matter\right)}\left(\zeta_{1}-\zeta_{1}'\right)$.
\begin{figure}
\begin{centering}
\includegraphics[scale=0.7]{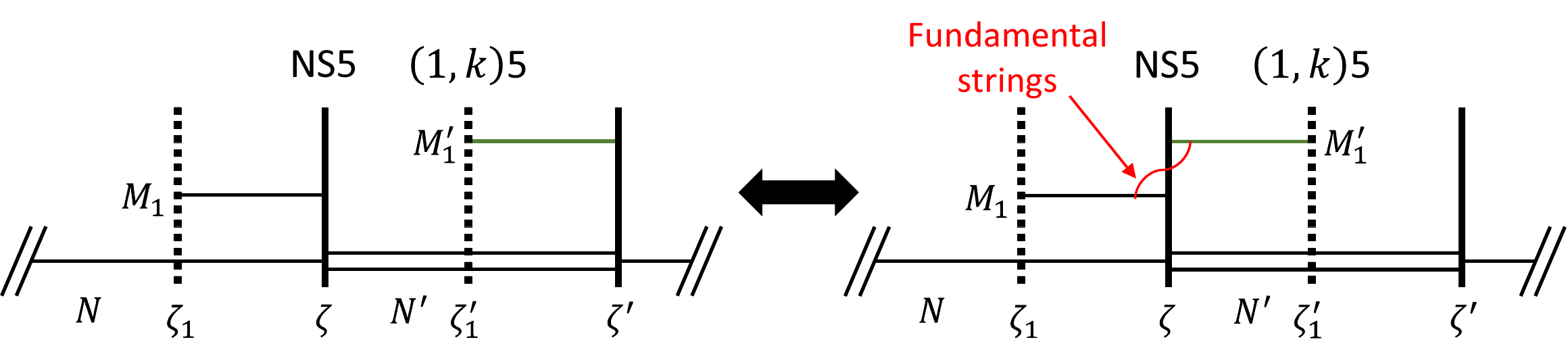}
\par\end{centering}
\caption{The brane transition in \eqref{eq:Duality-BC-22-ex}. The green D3-branes change the NS5-brane where they end before and after the transition. The red fundamental strings correspond to $Z_{M_{1},M_{1}'}^{\left(\protect\matter\right)}\left(\zeta_{1}-\zeta_{1}'\right)$ appearing in the right hand side of \eqref{eq:Duality-MM-22-ex} as discussed in appendix \ref{subsec:Factor-Meaning}. \label{fig:22-Trans-Ex}}
\end{figure}
The physical interpretation of this discrepancy is in figure \ref{fig:22-Trans-Ex}. By adding the decoupled sector to both sides, as a special case of \eqref{eq:CL-Dual-MM}, we obtain
\begin{equation}
Z_{\left(N,N+M_{1},N',N'+M_{1}'\right)}^{\left({\rm Left}\right)}Z_{\left(0,M_{1},N'-N+M_{1}',N'-N\right)}^{\left({\rm Right}\right)}=Z_{\left(N,N+M_{1},N'+M_{1}',N'\right)}^{\left({\rm Right}\right)}Z_{\left(0,M_{1},N'-N,N'-N+M_{1}'\right)}^{\left({\rm Left}\right)}.
\end{equation}
This implies the duality between
\begin{align}
 & \left[\mathrm{U}\left(N\right)_{-k}\times\mathrm{U}\left(N+M_{1}\right)_{k}\times\mathrm{U}\left(N'\right)_{-k}\times\mathrm{U}\left(N'+M_{1}'\right)_{k}\right]^{\mathrm{C}}\nonumber \\
 & \oplus\mathrm{U}\left(M_{1}\right)_{k}\times\mathrm{U}\left(N'-N+M_{1}'\right)_{-k}\times\mathrm{U}\left(N'-N\right)_{k},
\end{align}
and
\begin{align}
 & \left[\mathrm{U}\left(N\right)_{-k}\times\mathrm{U}\left(N+M_{1}\right)_{k}\times\mathrm{U}\left(N'+M_{1}'\right)_{-k}\times\mathrm{U}\left(N'\right)_{k}\right]^{\mathrm{C}}\nonumber \\
 & \oplus\mathrm{U}\left(M_{1}\right)_{k}\times\mathrm{U}\left(N'-N\right)_{-k}\times\mathrm{U}\left(N'-N+M_{1}'\right)_{k}.
\end{align}
This is a special case of the duality between \eqref{eq:BCdualL} and \eqref{eq:BCdualR}. Here the first line and the third line are the circular quiver diagrams, while the second line and the fourth line are the linear quiver diagram. The fourth line theory corresponds to the first line theory without the regular D3-branes, and the second line theory corresponds to the third line theory without the regular D3-branes.

Second, we consider the case of single NS5-brane obtained from \eqref{eq:Duality-BC-22-ex}. After performing the Hanany-Witten transition for the first $\left(1,k\right)$5-brane and the first NS5-brane and taking the cyclic rotation, the brane transition \eqref{eq:Duality-BC-22-ex} becomes
\begin{align}
 & \left\langle N'+M_{1}'\bullet_{\zeta'}N\bullet_{\zeta}N'+k-M_{1}\circ_{\zeta_{1}}N'\circ_{\zeta_{1}'}\right\rangle ^{\mathrm{P}}\nonumber \\
 & \leftrightarrow\left\langle N'\bullet_{\zeta'}N\bullet_{\zeta}N'+k-M_{1}+M_{1}'\circ_{\zeta_{1}}N'+M_{1}'\circ_{\zeta_{1}'}\right\rangle ^{\mathrm{P}}.
\end{align}
By identifying the two NS5-branes and relabeling $N'\rightarrow N$, $M_{1}\rightarrow k-M'$ and $M_{1}'=M$, we obtain
\begin{equation}
\left\langle N+M\bullet_{\zeta}N+M'\circ_{\zeta_{1}}N\circ_{\zeta_{1}'}\right\rangle ^{\mathrm{P}}\leftrightarrow\left\langle N\bullet_{\zeta}N+M+M'\circ_{\zeta_{1}}N+M\circ_{\zeta_{1}'}\right\rangle ^{\mathrm{P}}.\label{eq:Duality-BC-21-ex}
\end{equation}
This is the duality we presented in \eqref{eq:21Dual-Ex}. This duality is the typical case of \eqref{eq:Duality-NS1-BC}. The matrix model corresponding to the whole brane configuration \eqref{eq:BC-NS1-Diff} is
\begin{equation}
Z_{\left(\boldsymbol{N}\right)}=\int\frac{d^{N}\alpha}{N!}Z_{\bm{\zeta},\boldsymbol{N}}^{\left(-q,+1,-\bar{q}'\right)}\left(\alpha,\alpha\right).\label{eq:Zhole-Gen-Def-1}
\end{equation}
At the level of the matrix model, the brane transition \eqref{eq:Duality-NS1-BC} is realized as
\begin{equation}
Z_{\left(\boldsymbol{N}\right)}^{\left({\rm Left}\right)}Z_{\left(\bm{K}-N\bm{1}\right)}^{\left({\rm Right}\right)}=Z_{\left(\bm{K}\right)}^{\left({\rm Right}\right)}Z_{\left(\bm{N}-N\bm{1}\right)}^{\left({\rm Left}\right)},\label{eq:CL-Dual-MM-NS1}
\end{equation}
where $Z_{\left(\boldsymbol{N}\right)}^{\left({\rm Left}\right)}$ and $Z_{\left(\bm{K}\right)}^{\left({\rm Right}\right)}$ are the matrix models associated with the left side and the right side of \eqref{eq:Duality-NS1-BC}, respectively. The equality for \eqref{eq:Duality-BC-21-ex} is
\begin{equation}
Z_{\left(N+M_{1}',N+M_{1}+M_{1}',N\right)}^{\left({\rm Left}\right)}Z_{\left(M_{1},M_{1}'\right)}^{\left({\rm Right}\right)}=Z_{\left(N,N+M_{1},N'+M_{1}'\right)}^{\left({\rm Right}\right)}Z_{\left(M_{1}',M_{1}+M_{1}'\right)}^{\left({\rm Left}\right)}.\label{eq:Duality-MM-21-ex}
\end{equation}
This equality supports the conjectured duality \eqref{eq:21Dual-Ex}.

\subsection{Proof of relations among matrix models\label{subsec:Proof}}

In this section, we prove the relation \eqref{eq:Duality-MM}. To prove this relation, the Fermi gas-like approach plays important role. In the complete Fermi gas formalism, the matrix model is written as the partition function of an ideal Fermi gas with additional prefactors as explained in section \ref{subsec:QC}. However, this technique is still difficult to apply to general rank deformations. Therefore, we give up on applying it completely, and instead we focus on the advantages of this technique. In other words, we develop the computational techniques that are aligned with our objectives. We explain our improved Fermi gas approach in detail. For special cases, it has been known that the Fermi gas formalism can be applied for the local matrix model \cite{Kubo:2019ejc,Kubo:2020qed}. Our approach is based on this result. However, the numbers of D3-branes at the gluing points should be the same for applying the Fermi gas formalism completely. In this case, the numbers of D3-branes at both of the edges of the local brane configuration \eqref{eq:BCpart-Gen} should be the same. Here we extend the Fermi gas formalism so that we can apply it to the local matrix model \eqref{eq:MM-Gen} even when $N_{q+1}\neq\bar{N}_{\bar{q}+1}$. The computation is in appendix \ref{sec:FGF_loc}, and the result is \eqref{eq:FGF-Res}.

By using the result, we can rewrite \eqref{eq:Zhole-Gen-Def} as
\begin{align}
Z_{\left(\bm{\zeta},\bm{\zeta}'\right),\left(\boldsymbol{N},\bm{N}'\right)}= & \Omega_{\bm{\zeta},\bm{N}}^{q,\bar{q}}\Omega_{\bm{\zeta}',\bm{N}'}^{q',\bar{q}'}Z_{\left(\bm{M},\bar{\bm{M}}\right)}^{\left(\cstot\right)}Z_{\left(\bm{M}',\bar{\bm{M}}'\right)}^{\left(\cstot\right)}Z_{\left(\bm{\zeta}_{{\rm L}},\bm{\zeta}_{{\rm R}}\right),\left(\bm{M},\bar{\bm{M}}\right)}^{\left(\bifund\right)}Z_{\left(\bm{\zeta}_{{\rm L}}',\bm{\zeta}_{{\rm R}}'\right),\left(\bm{M}',\bar{\bm{M}}'\right)}^{\left(\bifund\right)}\nonumber \\
 & \times\int\frac{d^{N}\alpha}{\hbar^{N}N!}\frac{d^{N'}\beta}{\hbar^{N'}N'!}{\cal Z}_{\bm{\zeta},\boldsymbol{N}}^{q,\bar{q}}\left(\alpha,\beta\right){\cal Z}_{\bm{\zeta}',\boldsymbol{N}'}^{q',\bar{q}'}\left(\beta,\alpha\right).\label{eq:MM-FGF-gen}
\end{align}
This expression is useful for analyzing the behavior of the whole matrix models under the brane transition. It is important that the whole matrix model is factorized into four types of factors, which is the consequence of the result for the local matrix model in \eqref{eq:FGF-Res}. Let us check the behavior of each factor under the brane transition.

First of all, the two types of factors,
\begin{equation}
\Omega_{\bm{\zeta},\bm{N}}^{q,\bar{q}}\Omega_{\bm{\zeta}',\bm{N}'}^{q',\bar{q}'}\quad\text{and}\quad Z_{\left(\bm{\zeta}_{{\rm L}},\bm{\zeta}_{{\rm R}}\right),\left(\bm{M},\bar{\bm{M}}\right)}^{\left(\bifund\right)}Z_{\left(\bm{\zeta}_{{\rm L}}',\bm{\zeta}_{{\rm R}}'\right),\left(\bm{M}',\bar{\bm{M}}'\right)}^{\left(\bifund\right)},
\end{equation}
are not invariant under the brane transition. Therefore, these factors are removed in the exact relation \eqref{eq:Duality-MM}. However, surprisingly, the rest factors are invariant under the brane transition.

To see the invariance, we study the structure of each factor in detail. The Chern-Simons factor
\begin{equation}
Z_{\left(\bm{M},\bar{\bm{M}}\right)}^{\left(\cstot\right)}Z_{\left(\bm{M}',\bar{\bm{M}}'\right)}^{\left(\cstot\right)},
\end{equation}
is the product of the pure Chern-Simons factors $Z_{M}^{\left(\cs\right)}$ as shown in \eqref{eq:zCS-tot-Def}, where $M$ runs over all the numbers of the fractional D3-branes between an NS5-brane and a $\left(1,k\right)$5-brane. Namely, $M$ runs over $\left(\bm{M},\bar{\bm{M}},\bm{M}',\bar{\bm{M}}'\right)$. Because the set $\left(\bm{M},\bar{\bm{M}},\bm{M}',\bar{\bm{M}}'\right)$ is invariant before and after the brane transition (see \eqref{eq:Duality-BC-Set}), we conclude that this factor is invariant. The invariance of
\begin{equation}
{\cal Z}_{\bm{\zeta},\boldsymbol{N}}^{q,\bar{q}}\left(\alpha,\beta\right){\cal Z}_{\bm{\zeta}',\boldsymbol{N}'}^{q',\bar{q}'}\left(\beta,\alpha\right),\label{eq:FG-prod}
\end{equation}
is more obscure. It is important to notice that the factor ${\cal Z}_{\bm{\zeta},\boldsymbol{N}}^{q,\bar{q}}$ has the similar structure with $Z_{\left(\bm{M},\bar{\bm{M}}\right)}^{\left(\cstot\right)}$. Namely, ${\cal Z}_{\bm{\zeta},\boldsymbol{N}}^{q,\bar{q}}$ includes the products of
\begin{equation}
S_{\zeta,M}\left(\alpha\right)=\prod_{n}^{N}\frac{\prod_{j}^{M}2\sinh\frac{\alpha_{n}+2\pi\zeta+t_{M,j}}{2k}}{e^{\frac{\alpha_{n}+2\pi\zeta}{2}}+\left(-1\right)^{M_{a}}e^{-\frac{\alpha_{n}+2\pi\zeta}{2}}},\quad C_{\zeta,M}\left(\alpha\right)=\prod_{n}^{N}\frac{1}{\prod_{j}^{M}2\cosh\frac{\alpha_{n}+2\pi\zeta+t_{M,j}}{2k}},\label{eq:SC-Def}
\end{equation}
as shown in \eqref{eq:zFermi-Def}. The combination of $S_{\zeta,M}$ and $C_{\zeta,M}$ corresponds to one $\left(1,k\right)$5-brane $\circ_{\zeta,M}$ since $S_{\zeta,M}$ and $C_{\zeta,M}$ are parameterized by $\zeta$ and $M$. For ${\cal Z}_{\bm{\zeta},\boldsymbol{N}}^{q,\bar{q}}\left(\alpha,\beta\right)$, if a $\left(1,k\right)$5-brane $\circ_{\zeta,M}$ stands on the left side, it includes $S_{\zeta,M}\left(\alpha\right)C_{\zeta,M}\left(\beta\right)$ factor, while if a $\left(1,k\right)$5-brane $\circ_{\zeta,M}$ stands on the right side, it includes $C_{\zeta,M}\left(\alpha\right)S_{\zeta,M}\left(\beta\right)$. Remember that the transition changes the $\left(1,k\right)$5-branes standing on the left (right) side of an NS5-brane to the $\left(1,k\right)$5-branes standing on the right (left) side of the other NS5-brane. In \eqref{eq:FG-prod}, this transition corresponds to changing the $S_{\zeta,M}\left(\alpha\right)C_{\zeta,M}\left(\beta\right)$ ($C_{\zeta,M}\left(\alpha\right)S_{\zeta,M}\left(\beta\right)$) factor in ${\cal Z}_{\bm{\zeta},\boldsymbol{N}}^{q,\bar{q}}\left(\alpha,\beta\right)$ to the same factor in ${\cal Z}_{\bm{\zeta}',\boldsymbol{N}'}^{q',\bar{q}'}\left(\beta,\alpha\right)$. Therefore, under the brane transition, the product of them \eqref{eq:FG-prod} is also invariant.

Note that the invariance of \eqref{eq:FG-prod} strongly depends on their form. In other words, it is crucial that there are just two NS5-branes. However, in section \ref{sec:pq-model}, we will see that, for special cases, the similar brane transition exists for arbitrary even number of NS5-branes.

The simpler case \eqref{eq:CL-Dual-MM-NS1} can be proved much easier. The equality of the Chern-Simons factor is similarly obvious. The equality of the integral part
\begin{equation}
\int\frac{d^{N}\alpha}{\hbar^{N}N!}{\cal Z}_{\bm{\zeta},\boldsymbol{N}}^{q,\bar{q}}\left(\alpha,\alpha\right),
\end{equation}
is also obvious because ${\cal Z}_{\bm{\zeta},\boldsymbol{N}}^{q,\bar{q}}\left(\alpha,\alpha\right)$ includes the same factor $C_{\zeta,M}\left(\alpha\right)S_{\zeta,M}\left(\alpha\right)$ for $\circ_{\zeta,M}$ on both sides.

\section{Arbitrary even number of NS5-branes\label{sec:pq-model}}

In the previous section, we studied the (2,$q$) model (and the (1,$q$) model as a special case). In this section, we study (2$p$,$q$) model with integer $p$. Namely, we generalize the number of NS5-branes to arbitrary even numbers. We will show that similar relations exist when completely same $\left(1,k\right)$5-branes (namely, the number of D3-branes ending on and the FI parameter is the same) exist in every other segment between neighborhood NS5-branes. As in the previous section, we will start with describing dualities as brane transitions in section \ref{subsec:Duality-BC2}, and we will move to the detail of the dualities in section \ref{subsec:Duality-MM2}. The story is almost the same with the two NS5-branes case.

\subsection{Dualities as brane transitions\label{subsec:Duality-BC2}}

As explained in the previous section, when a brane configuration is a sequence of only the two local brane configurations including one NS5-brane \eqref{eq:BCpart-Gen} (which is the brane picture of the (2,$q$) model), there are non-trivial brane transitions. On the other hand, when a brane configuration is the sequence of more than the two local brane configurations \eqref{eq:BCpart-Gen}, in general, there are no similar dualities because the proof in section \ref{subsec:Proof} depends on the fact that there are only the two NS5-branes.

However, for special cases, accidental dualities appear. When the brane configuration is the sequence of the even number of local brane configurations \eqref{eq:BCpart-Gen}, and when every other local brane configuration has the same $\left(1,k\right)$5-branes in the same side, all of the D3-branes starting from the $\left(1,k\right)$5-branes can change the NS5-brane ending on to the opposite side simultaneously. Here, the same $\left(1,k\right)$5-branes means that both the number of D3-branes ending on the $\left(1,k\right)$5-branes and the shift parameter are the same.
\begin{figure}
\begin{centering}
\includegraphics[scale=0.7]{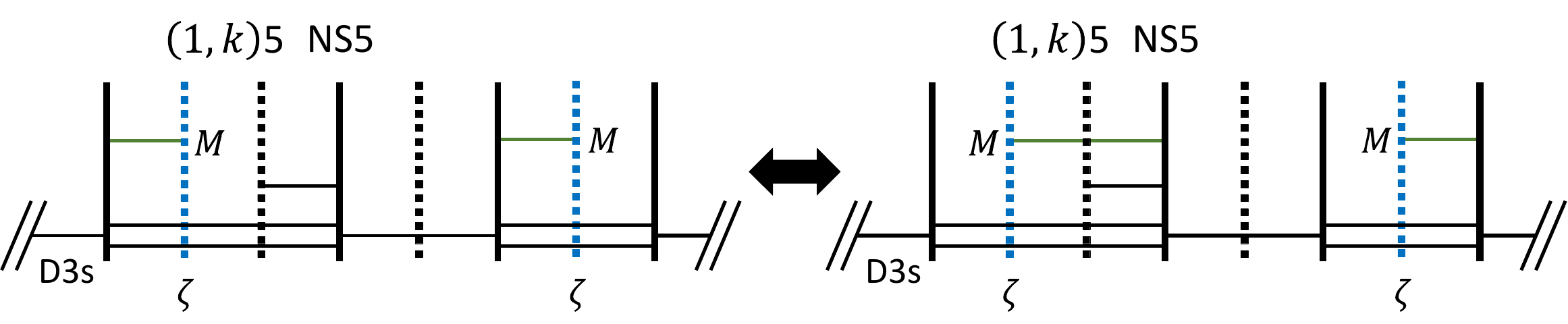}
\par\end{centering}
\caption{An example of the brane transition, where the brane configuration includes four NS5-branes. For every other segment between two NS5-branes, the same blue $\left(1,k\right)$5-brane exists. Namely, the shift parameters of them are the same $\zeta$, and the $M$ green fractional D3-branes end on them. The green fractional D3-branes simultaneously change the NS5-brane where they end on before and after the transition. All of the other branes are not changed before and after the transition.\label{fig:BC-trans2}}
\end{figure}
See figure \ref{fig:BC-trans2} as an example. The equivalent explanation is as follows. We start with the brane configuration
\begin{align}
\bigcup_{a=1}^{p} & \left(\left\langle N^{\left(2a-1\right)}\circ_{\zeta_{q^{\left(2a-1\right)}}^{\left(2a-1\right)},M_{q^{\left(2a-1\right)}}^{\left(2a-1\right)}}\cdots\circ_{\zeta_{1}^{\left(2a-1\right)},M_{1}^{\left(2a-1\right)}}\bullet_{\zeta^{\left(2a-1\right)}}\circ_{\bar{\zeta}_{1}^{\left(2a-1\right)},\bar{M}_{1}^{\left(2a-1\right)}}\cdots\circ_{\bar{\zeta}_{\bar{q}^{\left(2a-1\right)}}^{\left(2a-1\right)},\bar{M}_{\bar{q}^{\left(2a-1\right)}}^{\left(2a-1\right)}}N^{\left(2a\right)}\right\rangle \right.\nonumber \\
 & \quad\left.\cup\left\langle N^{\left(2a\right)}\circ_{\zeta_{q^{\left(2a\right)}}^{\left(2a\right)},M_{q^{\left(2a\right)}}^{\left(2a\right)}}\cdots\circ_{\zeta_{1}^{\left(2a\right)},M_{1}^{\left(2a\right)}}\bullet_{\zeta^{\left(2a\right)}}\circ_{\bar{\zeta}_{1}^{\left(2a\right)},\bar{M}_{1}^{\left(2a\right)}}\cdots\circ_{\bar{\zeta}_{\bar{q}^{\left(2a\right)}}^{\left(2a\right)},\bar{M}_{\bar{q}^{\left(2a\right)}}^{\left(2a\right)}}N^{\left(2a+1\right)}\right\rangle \right),\label{eq:pq-Dual-BC}
\end{align}
with the periodic boundary condition $N^{\left(2p+1\right)}=N^{\left(1\right)}$. (This is the brane picture of (2$p$,$Q$) model, where $Q=\sum_{a=1}^{2p}\left(q^{\left(a\right)}+\bar{q}^{\left(a\right)}\right)$.) Here $\cup$ means the gluing of the brane configurations. For example,
\begin{align}
 & \left\langle N\circ_{\zeta_{1},M_{1}}\bullet_{\zeta}\circ_{\bar{\zeta}_{1},\bar{M}_{1}}\circ N'\right\rangle \cup\left\langle \circ N'\circ_{\zeta_{1}',M_{1}'}\bullet_{\zeta'}\circ_{\bar{\zeta}_{1}',\bar{M}_{1}'}N''\right\rangle \nonumber \\
 & =\left\langle N\circ_{\zeta_{1},M_{1}}\bullet_{\zeta}\circ_{\bar{\zeta}_{1},\bar{M}_{1}}\circ N'\circ_{\zeta_{1}',M_{1}'}\bullet_{\zeta'}\circ_{\bar{\zeta}_{1}',\bar{M}_{1}'}N''\right\rangle .
\end{align}
We assume, for example, that $\bar{\zeta}_{\bar{q}^{\left(2a-1\right)}}^{\left(2a-1\right)}=\zeta$ and $\bar{M}_{\bar{q}^{\left(2a-1\right)}}^{\left(2a-1\right)}=M$ for all $a=1,2,\ldots,p$. In this special case, we find that all of the fractional D3-branes starting from the $\left(1,k\right)$5-branes $\circ_{\bar{\zeta}_{\bar{q}^{\left(2a-1\right)}}^{\left(2a-1\right)},\bar{M}_{\bar{q}^{\left(2a-1\right)}}^{\left(2a-1\right)}}$ can simultaneously change the NS5-branes where they end. Namely, the brane configuration \eqref{eq:pq-Dual-BC} is related to
\begin{align}
\bigcup_{a=1}^{p} & \left(\left\langle N^{\left(2a-1\right)}\circ_{\zeta_{q^{\left(2a-1\right)}}^{\left(2a-1\right)},M_{q^{\left(2a-1\right)}}^{\left(2a-1\right)}}\cdots\circ_{\zeta_{1}^{\left(2a-1\right)},M_{1}^{\left(2a-1\right)}}\right.\right.\nonumber \\
 & \quad\quad\quad\quad\quad\quad\quad\quad\quad\quad\quad\quad\quad\left.\bullet_{\zeta^{\left(2a-1\right)}}\circ_{\bar{\zeta}_{1}^{\left(2a-1\right)},\bar{M}_{1}^{\left(2a-1\right)}}\cdots\circ_{\bar{\zeta}_{\bar{q}^{\left(2a-1\right)}-1}^{\left(2a-1\right)},\bar{M}_{\bar{q}^{\left(2a-1\right)}-1}^{\left(2a-1\right)}}N^{\left(2a\right)}\right\rangle \nonumber \\
 & \quad\left.\cup\left\langle N^{\left(2a\right)}\circ_{\zeta,M}\circ_{\zeta_{q^{\left(2a\right)}}^{\left(2a\right)},M_{q^{\left(2a\right)}}^{\left(2a\right)}}\cdots\circ_{\zeta_{1}^{\left(2a\right)},M_{1}^{\left(2a\right)}}\bullet_{\zeta^{\left(2a\right)}}\circ_{\bar{\zeta}_{1}^{\left(2a\right)},\bar{M}_{1}^{\left(2a\right)}}\cdots\circ_{\bar{\zeta}_{\bar{q}^{\left(2a\right)}}^{\left(2a\right)},\bar{M}_{\bar{q}^{\left(2a\right)}}^{\left(2a\right)}}N^{\left(2a+1\right)}\right\rangle \right),
\end{align}
Here we assumed $M\leq M_{b^{\left(2a\right)}}^{\left(2a\right)}$ for all $a=1,2,\ldots,p$ and $b^{\left(2a\right)}=1,2,\ldots,q^{\left(2a\right)}$ just for the simplicity. If this condition is not satisfied, we have to move $\circ_{\zeta,M}$ to the correct place so that the ``good'' condition \eqref{eq:BCCond-Good} is satisfied.

Note that this transition includes the transition of the (2,$q$) models studied in section \ref{sec:2q-model} as a special case. For the (2,$q$) models, there are no restrictions for $\left(1,k\right)$5-branes since the brane configuration is the sequence of two local brane configurations including one NS5-brane \eqref{eq:BCpart-Gen}.

Since the explanation above is complicated, we give an example before closing this section. Let us consider the following brane configuration
\begin{equation}
\left\langle N_{2}+M\circ_{\zeta}N_{2}\bullet_{\eta_{2}}N_{3}\bullet_{\eta_{3}}N_{5}+M\circ_{\zeta}N_{5}\bullet_{\eta_{5}}N_{6}\bullet_{\eta_{6}}\right\rangle ^{\mathrm{P}}.
\end{equation}
This brane configuration is the combination of the four local brane configurations of the type of \eqref{eq:BCpart-Gen} as
\begin{equation}
\left\langle N_{6}\bullet_{\eta_{6}}\circ_{\zeta,M}N_{2}\right\rangle \cup\left\langle N_{2}\bullet_{\eta_{2}}N_{3}\right\rangle \cup\left\langle N_{3}\bullet_{\eta_{3}}\circ_{\zeta,M}N_{5}\right\rangle \cup\left\langle N_{5}\bullet_{\eta_{5}}N_{6}\right\rangle ,
\end{equation}
with the periodic boundary condition. Since every other local brane configuration has the same $\left(1,k\right)$5-brane $\circ_{\zeta,M}$, the D3-branes extended from the $\left(1,k\right)$5-brane $\circ_{\zeta,M}$ can simultaneously change the NS5-brane where they end as
\begin{equation}
\left\langle N_{6}\bullet_{\eta_{6}}N_{2}\right\rangle \cup\left\langle N_{2}\circ_{\zeta,M}\bullet_{\eta_{2}}N_{3}\right\rangle \cup\left\langle N_{3}\bullet_{\eta_{3}}N_{5}\right\rangle \cup\left\langle N_{5}\circ_{\zeta,M}\bullet_{\eta_{5}}N_{6}\right\rangle .
\end{equation}
In summary, we find the transition
\begin{align}
 & \left\langle N_{2}+M\circ_{\zeta}N_{2}\bullet_{\eta_{2}}N_{3}\bullet_{\eta_{3}}N_{5}+M\circ_{\zeta}N_{5}\bullet_{\eta_{5}}N_{6}\bullet_{\eta_{6}}\right\rangle ^{\mathrm{P}}\nonumber \\
 & \leftrightarrow\left\langle N_{2}\circ_{\zeta}N_{2}+M\bullet_{\eta_{2}}N_{3}\bullet_{\eta_{3}}N_{5}\circ_{\zeta}N_{5}+M\bullet_{\eta_{5}}N_{6}\bullet_{\eta_{6}}\right\rangle ^{\mathrm{P}}.\label{eq:BC-trans-Ex2}
\end{align}

\subsection{Dualities in quiver gauge theories\label{subsec:Duality-MM2}}

In this section, we study the dualities presented in the previous section in more detail by using the matrix model. The brane transitions are again realized in terms of the matrix model in the similar way as the (2,$q$) model. Namely, relations similar to \eqref{eq:Duality-MM} hold.

The matrix model is factorized into four factors by using \eqref{eq:FGF-Res} as
\begin{equation}
Z=\Omega Z^{\left(\cstot\right)}Z^{\left(\bifund\right)}Z^{\left(\fermi\right)},
\end{equation}
where $\Omega$ is the collection of $\Omega_{\bm{\zeta},\bm{N}}^{q,\bar{q}}$, $Z^{\left(\cstot\right)}$ is the collection of $Z_{\left(\bm{M},\bar{\bm{M}}\right)}^{\left(\cstot\right)}$, $Z^{\left(\bifund\right)}$ is the collection of $Z_{\left(\bm{\zeta}_{{\rm L}};\bm{\zeta}_{{\rm R}}\right),\left(\bm{M};\bar{\bm{M}}\right)}^{\left(\bifund\right)}$ and $Z^{\left(\fermi\right)}$ consists of the rest part. We find that, as with the (2,$q$) model case, $Z^{\left(\cstot\right)}$ and $Z^{\left(\fermi\right)}$ are the same in the dual theories while $\Omega$ and $Z^{\left(\bifund\right)}$ are different in general. The equality of $Z^{\left(\cstot\right)}$ is again trivial since this factor is the product of all of $M$ appearing in the whole brane configuration. Let us explain why $Z^{\left(\fermi\right)}$ is also invariant. The way of proof of the relation is similar to the (2,$q$) model case.

The explicit expression of this factor is
\begin{equation}
Z^{\left(\fermi\right)}=\int\left(\prod_{a}^{2p}d^{N^{\left(a\right)}}\lambda^{\left(a\right)}\right)\left(\prod_{a}^{2p}{\cal Z}_{\bm{\zeta}^{\left(a\right)},\boldsymbol{N}^{\left(a\right)}}^{q^{\left(a\right)},\bar{q}^{\left(a\right)}}\left(\lambda^{\left(a\right)},\lambda^{\left(a+1\right)}\right)\right),
\end{equation}
with the periodic condition $\lambda^{\left(2p+1\right)}=\lambda^{\left(1\right)}$. The important assumption is that, as explained in the previous section, every other local brane configuration has the same $\left(1,k\right)$5-branes $\circ_{\zeta,M}$ in the left (right) side. In terms of the matrix model, this means that all of the factor ${\cal Z}_{\bm{\zeta}^{\left(2a-1\right)},\boldsymbol{N}^{\left(2a-1\right)}}^{q^{\left(2a-1\right)},\bar{q}^{\left(2a-1\right)}}$ ($a=1,2,\ldots,p$) include the same factor $S_{\zeta,M}C_{\zeta,M}$ ($C_{\zeta,M}S_{\zeta,M}$) defined in \eqref{eq:SC-Def}. (Even when all of the factor ${\cal Z}_{\bm{\zeta}^{\left(2a\right)},\boldsymbol{N}^{\left(2a\right)}}^{q^{\left(2a\right)},\bar{q}^{\left(2a\right)}}$ include the same $\left(1,k\right)$5-brane factor, we can shift the label $a$ so that this case reduces to the above case.) These factors can be regarded as the factors of ${\cal Z}_{\bm{\zeta}^{\left(2a\right)},\boldsymbol{N}^{\left(2a\right)}}^{q^{\left(2a\right)},\bar{q}^{\left(2a\right)}}$ ($a=1,2,\ldots,p$) simultaneously. Namely, when all of ${\cal Z}_{\bm{\zeta}^{\left(2a-1\right)},\boldsymbol{N}^{\left(2a-1\right)}}^{q^{\left(2a-1\right)},\bar{q}^{\left(2a-1\right)}}$ include $S_{\zeta,M}C_{\zeta,M}$, we can rewrite as
\begin{align}
 & {\cal Z}_{\bm{\zeta}^{\left(2a-1\right)},\boldsymbol{N}^{\left(2a-1\right)}}^{q^{\left(2a-1\right)},\bar{q}^{\left(2a-1\right)}}\left(\mu,\nu\right){\cal Z}_{\bm{\zeta}^{\left(2a\right)},\boldsymbol{N}^{\left(2a\right)}}^{q^{\left(2a\right)},\bar{q}^{\left(2a\right)}}\left(\nu,\rho\right){\cal Z}_{\bm{\zeta}^{\left(2a+1\right)},\boldsymbol{N}^{\left(2a+1\right)}}^{q^{\left(2a+1\right)},\bar{q}^{\left(2a+1\right)}}\left(\rho,\sigma\right)\nonumber \\
 & =\frac{{\cal Z}_{\bm{\zeta}^{\left(2a-1\right)},\boldsymbol{N}^{\left(2a-1\right)}}^{q^{\left(2a-1\right)},\bar{q}^{\left(2a-1\right)}}\left(\mu,\nu\right)}{C_{\zeta,M}\left(\nu\right)}\left(C_{\zeta,M}\left(\nu\right){\cal Z}_{\bm{\zeta}^{\left(2a\right)},\boldsymbol{N}^{\left(2a\right)}}^{q^{\left(2a\right)},\bar{q}^{\left(2a\right)}}\left(\nu,\rho\right)S_{\zeta,M}\left(\rho\right)\right)\frac{{\cal Z}_{\bm{\zeta}^{\left(2a+1\right)},\boldsymbol{N}^{\left(2a+1\right)}}^{q^{\left(2a+1\right)},\bar{q}^{\left(2a+1\right)}}\left(\rho,\sigma\right)}{S_{\zeta,M}\left(\rho\right)},
\end{align}
and when they include $C_{\zeta,M}S_{\zeta,M}$, we can rewrite as
\begin{align}
 & {\cal Z}_{\bm{\zeta}^{\left(2a-1\right)},\boldsymbol{N}^{\left(2a-1\right)}}^{q^{\left(2a-1\right)},\bar{q}^{\left(2a-1\right)}}\left(\mu,\nu\right){\cal Z}_{\bm{\zeta}^{\left(2a\right)},\boldsymbol{N}^{\left(2a\right)}}^{q^{\left(2a\right)},\bar{q}^{\left(2a\right)}}\left(\nu,\rho\right){\cal Z}_{\bm{\zeta}^{\left(2a+1\right)},\boldsymbol{N}^{\left(2a+1\right)}}^{q^{\left(2a+1\right)},\bar{q}^{\left(2a+1\right)}}\left(\rho,\sigma\right)\nonumber \\
 & =\frac{{\cal Z}_{\bm{\zeta}^{\left(2a-1\right)},\boldsymbol{N}^{\left(2a-1\right)}}^{q^{\left(2a-1\right)},\bar{q}^{\left(2a-1\right)}}\left(\mu,\nu\right)}{S_{\zeta,M}\left(\nu\right)}\left(S_{\zeta,M}\left(\nu\right){\cal Z}_{\bm{\zeta}^{\left(2a\right)},\boldsymbol{N}^{\left(2a\right)}}^{q^{\left(2a\right)},\bar{q}^{\left(2a\right)}}\left(\nu,\rho\right)C_{\zeta,M}\left(\rho\right)\right)\frac{{\cal Z}_{\bm{\zeta}^{\left(2a+1\right)},\boldsymbol{N}^{\left(2a+1\right)}}^{q^{\left(2a+1\right)},\bar{q}^{\left(2a+1\right)}}\left(\rho,\sigma\right)}{C_{\zeta,M}\left(\rho\right)}.
\end{align}
This deformation explicitly corresponds to the brane transitions explained in the previous section. Therefore, $Z^{\left(\fermi\right)}$ is invariant under the brane transition. Note that this explanation shows the reason why we need the same $\left(1,k\right)$5-factor in every other factor ${\cal Z}_{\bm{\zeta}^{\left(a\right)},\boldsymbol{N}^{\left(a\right)}}^{q^{\left(a\right)},\bar{q}^{\left(a\right)}}$.

As a result, we obtain
\begin{equation}
Z_{\left(\boldsymbol{N},\bm{N}'\right)}^{\left({\rm Left}\right)}Z_{\left(\bm{K}-N_{{\rm min}}\bm{1},\bm{K}'-N_{{\rm min}}\bm{1}\right)}^{\left({\rm Right}\right)}=Z_{\left(\bm{K},\bm{K}'\right)}^{\left({\rm Right}\right)}Z_{\left(\bm{N}-N_{{\rm min}}\bm{1},\bm{N}'-N_{{\rm min}}\bm{1}\right)}^{\left({\rm Left}\right)},
\end{equation}
where $N_{{\rm min}}=\min_{a=1,2,\ldots,p}\left\{ N^{\left(a\right)}\right\} $. $Z_{\left(\boldsymbol{N},\bm{N}'\right)}^{\left({\rm Left}\right)}$ and $Z_{\left(\bm{K},\bm{K}'\right)}^{\left({\rm Right}\right)}$ are the matrix models associated to the dual brane configurations. This is the generalized version of \eqref{eq:CL-Dual-MM}. As with the (2,$q$) model, this identity implies the dualities in the theories which are the product of the circular quiver theory and the decoupled linear quiver theory.

Note that, when both of the dual matrix models satisfy the special property, $Z^{\left(\bifund\right)}=1$, we find
\begin{equation}
\left|Z_{\left(\boldsymbol{N},\bm{N}'\right)}^{\left({\rm Left}\right)}\right|=\left|Z_{\left(\bm{K},\bm{K}'\right)}^{\left({\rm Right}\right)}\right|.
\end{equation}
This relation suggests the non-trivial dualities in the circular quiver gauge theories. In other words, the dualities hold without the decoupled sector. An example of this case is \eqref{eq:BC-trans-Ex2}.

\section{Comparison with implications from quantum curve\label{sec:Res-QC}}

In this section, we compare our brane transition studied in the previous sections to \eqref{eq:ExcepSymD5}, \eqref{eq:ExcepSymE7} and \eqref{eq:FMN-trans}, which come form the symmetries of the quantum curves.

Before comparing, we remark about the region of the parameters. For the quantum curves, the parameter region is not restricted. On the other hand, for the brane configuration and the gauge theory, the parameter region is restricted so that the theory is ``good'' \eqref{eq:BCCond-Good} and the duality cascade does not occurs \eqref{eq:BCCond-SB}. Therefore, in the following we focus on the restricted region without mentioning it.

First, we find that our brane transitions include \eqref{eq:ExcepSymD5} and \eqref{eq:ExcepSymE7}. For example, \eqref{eq:Duality-BC-22-ex} is \eqref{eq:ExcepSymD5} with 
\begin{equation}
N_{1}=N,\quad N_{2}=N+M_{1},\quad N_{3}=N',\quad N_{4}=N'+M_{1}'.
\end{equation}
The first line of \eqref{eq:ExcepSymE7} comes from \eqref{eq:Duality-BC}. For example, when $N_{2}=N_{3}+M$ where $M\geq0$, our transition leads to
\begin{align}
 & \left\langle N_{1}\bullet_{\zeta}N_{3}+M\circ_{\zeta_{1}}N_{3}\bullet_{\zeta'}N_{4}\circ_{\zeta_{2}}N_{5}\circ_{\zeta_{3}}N_{6}\circ_{\zeta_{4}}\right\rangle ^{\mathrm{P}}\nonumber \\
 & \leftrightarrow\left\langle N_{1}\bullet_{\zeta}N_{3}\circ_{\zeta_{1}}N_{3}+M\bullet_{\zeta'}N_{4}\circ_{\zeta_{2}}N_{5}\circ_{\zeta_{3}}N_{6}\circ_{\zeta_{4}}\right\rangle ^{\mathrm{P}}.
\end{align}
This is the first line of \eqref{eq:ExcepSymE7}. Note that we do not have to impose the balanced condition \eqref{eq:Balance-Cond} for this transition.

On the other hand, the second line of \eqref{eq:ExcepSymE7} does not come from \eqref{eq:BC-trans-Ex2}. Instead, we have to use the brane transition for more general brane configuration studied in section \ref{sec:pq-model}. Here we have to replace the NS5-branes with the $\left(1,k\right)$5-branes and the $\left(1,k\right)$5-branes with the NS5-branes. Exchanging the two types of 5-branes corresponds to taking the complex conjugate to the matrix models as shown in \eqref{eq:MM-Conj}, and this operation does not affect our analysis for the matrix models. Therefore, after exchanging the two kind of 5-branes, \eqref{eq:BC-trans-Ex2} becomes
\begin{align}
 & \left\langle N_{2}+M\bullet_{\zeta}N_{2}\circ_{\eta_{2}}N_{3}\circ_{\eta_{3}}N_{5}+M\bullet_{\zeta}N_{5}\circ_{\eta_{5}}N_{6}\circ_{\eta_{6}}\right\rangle ^{\mathrm{P}}\nonumber \\
 & \leftrightarrow\left\langle N_{2}\bullet_{\zeta}N_{2}+M\circ_{\eta_{2}}N_{3}\circ_{\eta_{3}}N_{5}\bullet_{\zeta}N_{5}+M\circ_{\eta_{5}}N_{6}\circ_{\eta_{6}}\right\rangle ^{\mathrm{P}}.
\end{align}
This is the second line of \eqref{eq:ExcepSymE7} with
\begin{equation}
N_{1}=N_{2}+M,\quad N_{4}=N_{5}+M.
\end{equation}
Note that, in this case, the balanced condition \eqref{eq:Balance-Cond} is important because the balanced condition guarantees the difference between $N_{1}$ and $N_{2}$ and the difference between $N_{3}$ and $N_{4}$ are the same. As explained in section \ref{sec:pq-model}, this condition (and also the condition that the shift parameters $\zeta$ are the same) is essential for existing the transition for general brane configurations. We emphasize that, on the quantum curve side, the FI parameters are not considered, while our results include the FI parameters.

We finally comment on the local brane transition \eqref{eq:FMN-trans}. When the whole brane configuration consists of just two NS5-branes, we can bring the NS5-branes close by using the Hanany-Witten transition so that there are only one $\left(1,k\right)$5-brane between them. In this case, the transition \eqref{eq:Duality-BC} implies
\begin{align}
 & \left\langle N\circ_{\zeta_{q},M_{q}}\cdots\circ_{\zeta_{1},M_{1}}\bullet_{\zeta}\circ_{\bar{\zeta}_{1},\bar{M}_{1}}N'\bullet_{\zeta'}\circ_{\bar{\zeta}_{1}',\bar{M}_{1}'}\cdots\circ_{\bar{\zeta}_{\bar{q}'}',\bar{M}_{\bar{q}'}'}\right\rangle \nonumber \\
 & \leftrightarrow\left\langle N\circ_{\eta_{q},L_{q}}\cdots\circ_{\eta_{1},L_{1}}\bullet_{\zeta}N'\circ_{\bar{\zeta}_{1},\bar{M}_{1}}\bullet_{\zeta'}\circ_{\bar{\eta}_{1}',\bar{L}_{1}'}\cdots\circ_{\bar{\eta}_{\bar{q}'}',\bar{L}_{\bar{q}'}'}\right\rangle .
\end{align}
This is the local brane configuration. Therefore, the local brane configuration is equivalent to our result when the number of NS5-branes is two.

\section{Conclusion and discussion\label{sec:Discussion}}

In this paper, we studied the matrix model computing the round sphere partition function of three-dimensional $\mathcal{N}=4$ superconformal Chern-Simons theory with linear or circular quiver diagram. We find the functional relations among the matrix models by using the improved Fermi gas formalism. This result implies the dualities in the theories which are the product of the circular quiver theory and the decoupled linear quiver theory. We also interpret the dualities as the brane transitions in the Hanany-Witten setup. We found that the dualities can be quite simply understood in the brane picture. Since the brane transitions have been also conjectured from the symmetry of the quantum curve, we argued the relation between our result and the implication from the quantum curve. We found that our transitions include the transitions from the quantum curve.

Let us comment on the implications of our result to some topics related to the quantum curve. The improved Fermi gas formalism for the local theory \eqref{eq:FGF-Res} is consistent with the conjectured Fermi gas formalism \eqref{eq:MM-QC} because the prefactors in \eqref{eq:FGF-Res} are independent of the lowest rank. In addition, we showed in section \ref{sec:Res-QC} that our result and the implication from the quantum curve are consistent. Because the conjectured Fermi gas formalism and the conjectured parameter identification between $\bm{e}$ and $\bm{N}_{{\rm rel}}$ were used for obtaining the implication from the quantum curve \cite{Kubo:2019ejc}, our result supports the conjectured Fermi gas formalism and the parameter identification.

We shall list some future directions in the following. First, it is interesting to extend our result to other $\mathcal{N}=4$ superconformal Chern-Simons theories. Especially, the Fermi gas formalism for the affine D-type quiver theory \cite{Assel:2015hsa,Moriyama:2015jsa} and the ${\rm O}\left(2N+1\right)\times{\rm USp}\left(2N+2M\right)$ gauge theory \cite{Okuyama:2016xke,Moriyama:2016xin} is known, and thus there is possibility to apply our result to these theories.

Second, it is important to provide other checks of the conjectured dualities, for example, by studying the loop operators or superconformal indices. The supersymmetric localization technique would be helpful also for these cases. It would be also important to find some connections between the conjectured dualities and some known dualities. There might be a possibility that the conjectured dualities are supported by lifting the brane construction to M-theory.

Third, it is known that the spectral determinant of the quantum curve or the Fermi gas partition function is related to integrable systems such that the $q$-Painlev\'e equations or the affine ${\rm SU}\left(\nu\right)$ $q$-Toda equations \cite{Bonelli:2017gdk,Nosaka:2020tyv}. The relation among the matrix models which we found would be closely related to the symmetry of these integrable systems. This is because the Weyl group symmetry of the quantum curve, which is the strong motivation of our study, is related to the affine Weyl symmetry of $q$-Painlev\'e equation \cite{Moriyama:2021mux}. However, the role of the prefactors independent of the lowest rank in the context of the integrability has been not obvious yet. Because we also studied the prefactors in detail, we hope that our study will be helpful for revealing the integrability of the whole matrix models and, moreover, the $\mathcal{N}=4$ superconformal Chern-Simons theories.

\section*{Acknowledgments}

We are grateful to Masazumi Honda, Keita Nii, Seiji Terashima, Shuichi Yokoyama and especially Sanefumi Moriyama and Tomoki Nosaka for valuable discussions and comments. This work was supported by Grant-in-Aid for JSPS Fellows No.20J12263.

\appendix

\section{Fermi gas approach for local matrix model\label{sec:FGF_loc}}

In this appendix we derive the Fermi gas formalism for the local matrix model \eqref{eq:MM-Gen}, which corresponds to local brane configuration including one NS5-brane \eqref{eq:BCpart-Gen}. We use the difference notation \eqref{eq:DiffNot} and assume that the brane configuration satisfies \eqref{eq:BCCond-Good}. The procedure for applying the Fermi gas formalism follows the work in \cite{Kubo:2020qed}. However, there are two differences. First, we do not assume that the number of D3-branes in both of the sides is the same. This is crucial for dealing with the general rank deformations. Second, we turn on the FI terms.

The identity which we will show is
\begin{align}
Z_{\bm{\zeta},\boldsymbol{N}}^{\left(-q,+1,-\bar{q}\right)}\left(\frac{\alpha}{\hbar},\frac{\beta}{\hbar}\right)= & \Omega_{\bm{\zeta},\bm{N}}^{q,\bar{q}}Z_{\left(\bm{M},\bar{\bm{M}}\right)}^{\left(\cstot\right)}Z_{\left(\bm{\zeta}_{{\rm L}};\bm{\zeta}_{{\rm R}}\right),\left(\bm{M};\bar{\bm{M}}\right)}^{\left(\bifund\right)}\nonumber \\
 & \times\int d^{N_{q+1}}\lambda^{\left(q+1\right)}d^{\bar{N}_{\bar{q}+1}}\bar{\lambda}^{\left(\bar{q}+1\right)}\prod_{n}^{N_{q+1}}\braket{\alpha_{n}|e^{\frac{i}{2\hbar}\hat{p}^{2}}|\lambda_{n}^{\left(q+1\right)}}\nonumber \\
 & \times{\cal Z}_{\bm{\zeta},\boldsymbol{N}}^{q,\bar{q}}\left(\lambda^{\left(q+1\right)},\bar{\lambda}^{\left(\bar{q}+1\right)}\right)\prod_{n}^{\bar{N}_{\bar{q}+1}}\braket{\lambda_{n}^{\left(\bar{q}+1\right)}|e^{-\frac{i}{2\hbar}\hat{p}^{2}}|\beta_{n}},\label{eq:FGF-Res}
\end{align}
where
\begin{align}
\bm{\zeta}_{{\rm L}} & =\left(\zeta_{1},\zeta_{2},\ldots,\zeta_{q}\right),\quad\bm{\zeta}_{{\rm R}}=\left(\bar{\zeta}_{1},\bar{\zeta}_{2},\ldots,\bar{\zeta}_{\bar{q}}\right),\nonumber \\
\bm{M} & =\left(M_{1},M_{2},\ldots,M_{q}\right),\quad\bar{\bm{M}}=\left(\bar{M}_{1},\bar{M}_{2},\ldots,\bar{M}_{q}\right).
\end{align}
Note that $e^{\pm\frac{i}{2\hbar}\hat{p}^{2}}$ factors in the second and third line vanish when we glue the theories because these factors in the second and third line are the inverses each other. The most important point here is that the matrix model factorizes into four kinds of factors. This is reminiscent of the matrix model in the Fermi gas formalism \eqref{eq:MM-QC}. We call $Z_{\left(\bm{M},\bar{\bm{M}}\right)}^{\left(\cstot\right)}$ and $Z_{\left(\bm{\zeta}_{{\rm L}};\bm{\zeta}_{{\rm R}}\right),\left(\bm{M};\bar{\bm{M}}\right)}^{\left(\bifund\right)}$ prefactors.

The definition of each factor is as follows. First,
\begin{equation}
\Omega_{\bm{\zeta},\bm{N}}^{q,\bar{q}}=e^{i\Theta_{\bm{\zeta},\bm{N}}^{q,\bar{q}}}e^{\frac{2\pi i\zeta}{k}\left(\sum_{a}^{q}M_{a}\zeta_{a}-\sum_{a}^{\bar{q}}\bar{M}_{a}\bar{\zeta}_{a}\right)},\label{eq:localMM-Phase-Def}
\end{equation}
is the phase factor, where
\begin{align}
\Theta_{\bm{\zeta},\boldsymbol{N}}^{q,\bar{q}} & =\sum_{a}^{q}\theta_{\zeta_{a},N_{a}-N_{a+1}}+\theta_{\zeta,N_{1}-\bar{N}_{1}}-\sum_{a}^{\bar{q}}\theta_{\bar{\zeta}_{a},\bar{N}_{a}-\bar{N}_{a+1}},\nonumber \\
\theta_{\zeta,M} & =\frac{\pi}{k}\left[\frac{1}{12}\left(M^{3}-M\right)-M\zeta^{2}\right].\label{eq:Framing-Def}
\end{align}
Second,
\begin{equation}
Z_{\left(\bm{M},\bar{\bm{M}}\right)}^{\left(\cstot\right)}=\prod_{a}^{q}Z_{M_{a}}^{\left(\cs\right)}\prod_{a}^{\bar{q}}Z_{\bar{M}_{a}}^{\left(\cs\right)},\label{eq:zCS-tot-Def}
\end{equation}
where
\begin{align}
Z_{M}^{\left(\cs\right)} & =\frac{1}{k^{\frac{M}{2}}}\prod_{j<j'}^{M}2\sin\frac{\pi}{k}\left(j'-j\right),\label{eq:zCS-Def}
\end{align}
is the partition function of ${\rm U}\left(M\right)_{k}$ pure Chern-Simons theory. Third,
\begin{equation}
Z_{\left(\bm{\zeta}_{{\rm L}};\bm{\zeta}_{{\rm R}}\right),\left(\bm{M};\bar{\bm{M}}\right)}^{\left(\bifund\right)}=\prod_{a<b}^{q}Z_{M_{a},M_{b}}^{\left(\vector\right)}\left(\zeta_{a}-\zeta_{b}\right)\prod_{a,b}^{q,\bar{q}}Z_{M_{a},\bar{M}_{b}}^{\left(\matter\right)}\left(\zeta_{a}-\bar{\zeta}_{b}\right)\prod_{a<b}^{\bar{q}}Z_{\bar{M}_{a},\bar{M}_{b}}^{\left(\vector\right)}\left(\bar{\zeta}_{a}-\bar{\zeta}_{b}\right),
\end{equation}
where
\begin{align}
Z_{M,M'}^{\left(\vector\right)}\left(\zeta\right) & =\prod_{j'}^{M'}\frac{\prod_{j}^{M}2\sinh\frac{1}{2k}\left(t_{M,j}-t_{M',j'}+\zeta\right)}{e^{\pi\zeta}-\left(-1\right)^{M+M'}e^{-\pi\zeta}},\nonumber \\
Z_{M,\bar{M}}^{\left(\matter\right)}\left(\zeta\right) & =\prod_{j}^{M}\prod_{\overline{j}}^{\bar{M}}\frac{1}{2\cosh\frac{1}{2k}\left(t_{M,j}-t_{\bar{M},\bar{j}}+\zeta\right)},\label{eq:Zvecmat-Def}
\end{align}
and
\begin{equation}
t_{M,j}=2\pi i\left(\frac{M+1}{2}-j\right).\label{eq:tM-Def}
\end{equation}
$Z_{M,M'}^{\left(\vector\right)}\left(\zeta\right)$ and $Z_{M,\bar{M}}^{\left(\matter\right)}\left(\zeta\right)$ are real functions because for each factor, there is one factor which is complex conjugate of it (namely, $j$ and $M-j$). These prefactors capture the dynamics of fundamental strings on the fractional D3-branes. In appendix \ref{subsec:Factor-Meaning}, we will explain this statement in detail. The prefactors \eqref{eq:localMM-Phase-Def}, \eqref{eq:zCS-tot-Def} and \eqref{eq:Zvecmat-Def} depend only on the relative ranks. In other words, all of $Z_{\bm{\zeta},\boldsymbol{N}+N\boldsymbol{1}}^{\left(-q,+1,-\bar{q}\right)}$ have the same prefactors for arbitrary $N$. Fourth,
\begin{align}
{\cal Z}_{\bm{\zeta},\boldsymbol{N}}^{q,\bar{q}}\left(\mu,\nu\right)= & S_{\bm{\zeta}_{{\rm L}},\bm{M}}\left(\mu\right)C_{\bm{\zeta}_{{\rm R}},\bar{\bm{M}}}\left(\mu\right)S_{\bm{\zeta}_{{\rm R}},\bar{\bm{M}}}\left(\nu\right)C_{\bm{\zeta}_{{\rm L}},\bm{M}}\left(\nu\right)\nonumber \\
 & \times e^{-\frac{i\zeta}{k}\left(\sum_{m}^{N_{q+1}}\mu_{m}-\sum_{n}^{\bar{N}_{\bar{q}+1}}\nu_{n}\right)}\frac{\prod_{m<m'}^{N_{q+1}}2\sinh\frac{\mu_{m}-\mu_{m'}}{2k}\prod_{n<n'}^{\bar{N}_{\bar{q}+1}}2\sinh\frac{\nu_{n}-\nu_{n'}}{2k}}{\prod_{m}^{N_{q+1}}\prod_{n}^{\bar{N}_{\bar{q}+1}}2\cosh\frac{\mu_{m}-\nu_{n}}{2k}},\label{eq:zFermi-Def}
\end{align}
where
\begin{align}
S_{\left(\zeta_{1},\zeta_{2},\ldots,\zeta_{r}\right),\left(M_{1},M_{2},\ldots,M_{r}\right)}\left(\mu_{1},\mu_{2},\ldots,\mu_{N}\right) & =\prod_{a}^{r}\prod_{n}^{N}\frac{\prod_{j}^{M_{a}}2\sinh\frac{\mu_{n}+2\pi\zeta_{a}+t_{M_{a},j}}{2k}}{e^{\frac{\mu_{n}+2\pi\zeta_{a}}{2}}+\left(-1\right)^{M_{a}}e^{-\frac{\mu_{n}+2\pi\zeta_{a}}{2}}},\nonumber \\
C_{\left(\zeta_{1},\zeta_{2},\ldots,\zeta_{r}\right),\left(M_{1},M_{2},\ldots,M_{r}\right)}\left(\mu_{1},\mu_{2},\ldots,\mu_{N}\right) & =\prod_{a}^{r}\prod_{n}^{N}\frac{1}{\prod_{j}^{M_{a}}2\cosh\frac{\mu_{n}+2\pi\zeta_{a}+t_{M_{a},j}}{2k}}.
\end{align}
This factor would include complete information of the Fermi gas partition function \eqref{eq:FGpart}.

\subsection{Derivation\label{subsec:Derivation}}

We start with the local matrix model \eqref{eq:MM-Gen}
\begin{align}
Z_{\bm{\zeta},\boldsymbol{N}}^{\left(-q,+1,-\bar{q}\right)}\left(\frac{\alpha}{\hbar},\frac{\beta}{\hbar}\right)= & \int\left(\prod_{a}^{q}\frac{d^{N_{a}}\lambda^{\left(a\right)}}{N_{a}!\hbar^{N_{a}}}\right)\left(\prod_{a}^{\bar{q}}\frac{d^{\bar{N}_{a}}\bar{\lambda}^{\left(a\right)}}{\bar{N}_{a}!\hbar^{\bar{N}_{a}}}\right)\prod_{a}^{q}Z_{\zeta_{a}\left(N_{a+1},N_{a}\right)}^{\left(-1\right)}\left(\frac{\lambda^{\left(a+1\right)}}{\hbar},\frac{\lambda^{\left(a\right)}}{\hbar}\right)\nonumber \\
 & \times Z_{\zeta,\left(N_{1},\bar{N}_{1}\right)}^{\left(+1\right)}\left(\frac{\lambda^{\left(1\right)}}{\hbar},\frac{\bar{\lambda}^{\left(1\right)}}{\hbar}\right)\prod_{a}^{\bar{q}}Z_{\bar{\zeta}_{a},\left(\bar{N}_{a},\bar{N}_{a+1}\right)}^{\left(-1\right)}\left(\frac{\bar{\lambda}^{\left(a\right)}}{\hbar},\frac{\bar{\lambda}^{\left(a+1\right)}}{\hbar}\right),\label{eq:MMgen2}
\end{align}
where we rescaled all of the integration variable as $\lambda\rightarrow\frac{\lambda}{\hbar}$.

We first use the determinant formula \cite{Honda:2020uou}
\begin{equation}
e^{-2\pi i\zeta\left(\sum_{n}^{N_{1}}\alpha_{n}-\sum_{n}^{N_{2}}\beta_{n}\right)}Z_{N_{1},N_{2}}^{\lp}\left(\frac{\alpha}{\hbar},\frac{\beta}{\hbar}\right)=\begin{cases}
\det\left(\begin{array}{c}
\left[F_{m,n}\right]_{m,n}^{N_{1}\times N_{2}}\\
\left[M_{j,n}^{{\rm L}}\right]_{j,n}^{\left(N_{2}-N_{1}\right)\times N_{2}}
\end{array}\right) & \left(N_{1}\leq N_{2}\right)\\
\det\left(\begin{array}{cc}
\left[F_{m,n}\right]_{m,n}^{N_{1}\times N_{2}} & \left[M_{m,j}^{{\rm R}}\right]_{m,j}^{N_{1}\times\left(N_{1}-N_{2}\right)}\end{array}\right) & \left(N_{1}>N_{2}\right)
\end{cases},\label{eq:CauchyDet}
\end{equation}
where
\begin{align}
F_{m,n} & =\hbar\braket{\alpha_{m}|\frac{1}{2\cosh\frac{\hat{p}+2\pi\zeta+i\pi\left(N_{1}-N_{2}\right)}{2}}|\beta_{n}},\nonumber \\
M_{j,n}^{{\rm L}} & =\frac{\hbar}{\sqrt{k}}\bbraket{t_{M,j}-2\pi\zeta|\beta_{n}},M_{m,j}^{{\rm R}}\quad=\frac{\hbar}{\sqrt{k}}\brakket{\alpha_{m}|-t_{M,j}-2\pi\zeta}.
\end{align}
We can apply this formula to $Z_{\zeta,\left(N_{1},N_{2}\right)}^{\left(\pm1\right)}\left(\alpha,\beta\right)$. The Fresnel factors remained in the $\left(1,k\right)$5-brane factors can be put between the bras and the kets. We then insert
\begin{equation}
1=\int d^{N_{q+1}}\lambda^{\left(q+1\right)}\prod_{n}^{N_{q+1}}\braket{\alpha_{n}|\lambda_{n}^{\left(q+1\right)}},\quad1=\int d^{\bar{N}_{\bar{q}+1}}\bar{\lambda}^{\left(\bar{q}+1\right)}\prod_{n}^{\bar{N}_{\bar{q}+1}}\braket{\bar{\lambda}_{n}^{\left(\bar{q}+1\right)}|\beta_{n}}.
\end{equation}
Now the matrix model is invariant under the following change
\begin{equation}
\ket q\bra q\rightarrow e^{\frac{i}{2\hbar}\hat{p}^{2}}\ket q\bra qe^{-\frac{i}{2\hbar}\hat{p}^{2}},
\end{equation}
for all the bras and kets. The NS5-brane factor does not change up to phases. On the other hand, the $\left(1,k\right)$5-brane factors drastically change. Momentum operators become position operators using
\begin{equation}
e^{-\frac{i}{2\hbar}\hat{p}^{2}}e^{-\frac{i}{2\hbar}\hat{q}^{2}}f\left(\hat{p}\right)e^{\frac{i}{2\hbar}\hat{q}^{2}}e^{\frac{i}{2\hbar}\hat{p}^{2}}=f\left(\hat{q}\right),
\end{equation}
and momentum eigenvectors become position eigenvectors using
\begin{equation}
e^{-\frac{i}{2\hbar}\hat{p}^{2}}e^{-\frac{i}{2\hbar}\hat{q}^{2}}\kket p=\frac{1}{\sqrt{i}}e^{\frac{i}{2\hbar}p^{2}}\ket p,\quad\bbra pe^{\frac{i}{2\hbar}\hat{q}^{2}}e^{\frac{i}{2\hbar}\hat{p}^{2}}=\sqrt{i}e^{-\frac{i}{2\hbar}p^{2}}\bra p.
\end{equation}
In this step, there appear two types of phase, namely $i^{\frac{1}{2}\left(N_{1}-N_{q+1}-\bar{N}_{1}+\bar{N}_{\bar{q}+1}\right)}$ and $e^{i\Theta_{\bm{\zeta},\bm{N}}^{q,\bar{q}}}$, which is defined in \eqref{eq:Framing-Def}. We obtain
\begin{align}
Z_{\bm{\zeta},\boldsymbol{N}}^{\left(-q,+1,-\bar{q}\right)}\left(\frac{\alpha}{\hbar},\frac{\beta}{\hbar}\right)= & i^{\frac{1}{2}\left(N_{q+1}^{2}-N_{1}^{2}+\bar{N}_{1}^{2}-\bar{N}_{\bar{q}+1}^{2}\right)+\frac{1}{2}\left(N_{1}-N_{q+1}-\bar{N}_{1}+\bar{N}_{\bar{q}+1}\right)}e^{i\Theta_{\bm{\zeta},\bm{N}}^{q,\bar{q}}}\nonumber \\
 & \times\int d^{N_{q+1}}\lambda^{\left(q+1\right)}d^{\bar{N}_{\bar{q}+1}}\bar{\lambda}^{\left(\bar{q}+1\right)}\prod_{n}^{N_{q+1}}\braket{\alpha_{n}|e^{\frac{i}{2\hbar}\hat{p}^{2}}|\lambda_{n}^{\left(q+1\right)}}\nonumber \\
 & \times\tilde{Z}_{\bm{\zeta},\boldsymbol{N}}^{\left(-q,+1,-\bar{q}\right)}\left(\frac{\lambda^{\left(q+1\right)}}{\hbar},\frac{\bar{\lambda}^{\left(\bar{q}+1\right)}}{\hbar}\right)\prod_{n}^{\bar{N}_{\bar{q}+1}}\braket{\lambda_{n}^{\left(\bar{q}+1\right)}|e^{-\frac{i}{2\hbar}\hat{p}^{2}}|\beta_{n}},\label{eq:Zcomp2}
\end{align}
where
\begin{align}
 & \tilde{Z}_{\bm{\zeta},\boldsymbol{N}}^{\left(-q,+1,-\bar{q}\right)}\left(\frac{\lambda^{\left(q+1\right)}}{\hbar},\frac{\bar{\lambda}^{\left(\bar{q}+1\right)}}{\hbar}\right)\nonumber \\
 & =\int\prod_{a}^{q}\frac{d^{N_{a}}\lambda^{\left(a\right)}}{N_{a}!\hbar^{N_{a}}}\prod_{a}^{\bar{q}}\frac{d^{\bar{N}_{a}}\bar{\lambda}^{\left(a\right)}}{\bar{N}_{a}!\hbar^{\bar{N}_{a}}}\prod_{a}^{q}\det\left(\begin{array}{c}
\left[\hbar\bra{\lambda_{m}^{\left(a+1\right)}}\frac{1}{2\cosh\frac{\hat{q}+2\pi\zeta_{a}-i\pi M_{a}}{2}}\ket{\lambda_{n}^{\left(a\right)}}\right]_{m,n}^{N_{a+1}\times N_{a}}\\
\left[\frac{\hbar}{\sqrt{k}}\braket{t_{M_{a},j}-2\pi\zeta_{a}|\lambda_{n}^{\left(a\right)}}\right]_{j,n}^{M_{a}\times N_{a}}
\end{array}\right)\nonumber \\
 & \quad\times Z_{\xi,\left(N_{1},\bar{N}_{1}\right)}^{\left(+1\right)}\left(\frac{\lambda^{\left(1\right)}}{\hbar},\frac{\bar{\lambda}^{\left(1\right)}}{\hbar}\right)\nonumber \\
 & \quad\times\prod_{a}^{\bar{q}}\det\left(\begin{array}{cc}
\left[\hbar\bra{\bar{\lambda}_{m}^{\left(a\right)}}\frac{1}{2\cosh\frac{\hat{q}+2\pi\bar{\zeta}_{a}+i\pi\bar{M}_{a}}{2}}\ket{\bar{\lambda}_{n}^{\left(a+1\right)}}\right]_{m,n}^{\bar{N}_{a}\times\bar{N}_{a+1}} & \left[\frac{\hbar}{\sqrt{k}}\braket{\bar{\lambda}_{m}^{\left(a\right)}|-t_{\bar{M}_{a},j}-2\pi\bar{\zeta}_{a}}\right]_{m,j}^{\bar{N}_{a}\times\bar{M}_{a}}\end{array}\right).
\end{align}
Since the NS5-brane factor $Z_{\xi,\left(N_{1},\bar{N}_{1}\right)}^{\left(+1\right)}$ and all of the determinants are anti-symmetric functions, we can diagonalize all the determinants. We apply the formula
\begin{align}
\frac{1}{N!}\int d^{N}\lambda f\left(\lambda_{1},\lambda_{2},\ldots,\lambda_{N}\right)\det\left(\left[g_{n}\left(\lambda_{m}\right)\right]_{m,n}^{N\times N}\right)=\int d^{N}\lambda f\left(\lambda_{1},\lambda_{2},\ldots,\lambda_{N}\right)\prod_{m}^{N}g_{m}\left(\lambda_{m}\right),
\end{align}
which holds for any anti-symmetric function $f\left(\bm{\lambda}\right)$, to all of the determinants. Because the inner products of position vectors become delta functions, we can perform all of the integrations.\footnote{To perform the integration by using the delta function with complex number argument, we have to shift the integration contour to the imaginary direction. The ``good'' condition \eqref{eq:BCCond-Good} and the ``no duality cascade'' condition \eqref{eq:BCCond-SB} guarantee that the contour does not pass through any poles \cite{Assel:2014awa,Kubo:2020qed}.} We obtain
\begin{align}
\tilde{Z}_{\bm{\zeta},\boldsymbol{N}}^{\left(-q,+1,-\bar{q}\right)}\left(\frac{\mu}{\hbar},\frac{\nu}{\hbar}\right)= & k^{-\frac{N_{1}-N_{q+1}}{2}-\frac{\bar{N}_{1}-\bar{N}_{\bar{q}+1}}{2}}\prod_{a}^{q}\prod_{n_{a+1}}^{N_{a+1}}\frac{1}{2\cosh\frac{\lambda_{n_{a+1}}-i\pi M_{a}+2\pi\zeta_{a}}{2}}\nonumber \\
 & \times Z_{\zeta,\left(N_{1},\bar{N}_{1}\right)}^{\left(+1\right)}\left(\frac{\lambda}{\hbar},\frac{\bar{\lambda}}{\hbar}\right)\prod_{a}^{\bar{q}}\prod_{n_{a+1}}^{\bar{N}_{a+1}}\frac{1}{2\cosh\frac{\bar{\lambda}_{n_{a+1}}+i\pi\bar{M}_{a}+2\pi\bar{\zeta}_{a}}{2}},\label{eq:Zcomp3}
\end{align}
where
\begin{equation}
\lambda_{n}=\begin{cases}
\mu_{n} & \left(1\leq n\leq N_{q+1}\right)\\
t_{M_{q},n-N_{q+1}}-2\pi\zeta_{q} & \left(N_{q+1}+1\leq n\leq N_{q}\right)\\
t_{M_{q-1},n-N_{q}}-2\pi\zeta_{q-1} & \left(N_{q}+1\leq n\leq N_{q-1}\right)\\
\,\,\,\,\,\,\,\,\vdots\\
t_{M_{1},n-N_{2}}-2\pi\zeta_{1} & \left(N_{2}+1\leq n\leq N_{1}\right)
\end{cases},
\end{equation}
and
\begin{equation}
\bar{\lambda}_{n}=\begin{cases}
\nu_{n} & \left(1\leq n\leq\bar{N}_{\bar{q}+1}\right)\\
-t_{\bar{M}_{\bar{q}},n-\bar{N}_{\bar{q}+1}}-2\pi\bar{\zeta}_{\bar{q}} & \left(\bar{N}_{\bar{q}+1}+1\leq n\leq\bar{N}_{\bar{q}}\right)\\
-t_{\bar{M}_{\bar{q}-1},n-\bar{N}_{\bar{q}}}-2\pi\bar{\zeta}_{\bar{q}-1} & \left(\bar{N}_{\bar{q}}+1\leq n\leq\bar{N}_{\bar{q}-1}\right)\\
\,\,\,\,\,\,\,\,\vdots\\
-t_{\bar{M}_{1},n-\bar{N}_{2}}-2\pi\bar{\zeta}_{1} & \left(\bar{N}_{2}+1\leq n\leq\bar{N}_{1}\right)
\end{cases}.
\end{equation}
The NS5-brane factor $Z_{\zeta,\left(N_{1},\bar{N}_{1}\right)}^{\left(+1\right)}$ becomes
\begin{align}
 & k^{-\frac{N_{1}-N_{q+1}}{2}-\frac{\bar{N}_{1}-\bar{N}_{\bar{q}+1}}{2}}Z_{\zeta,\left(N_{1},\bar{N}_{1}\right)}^{\left(+1\right)}\left(\frac{\lambda}{\hbar},\frac{\bar{\lambda}}{\hbar}\right)\nonumber \\
 & =e^{\frac{2\pi i\zeta}{k}\left(\sum_{a}^{q}M_{a}\zeta_{a}-\sum_{a}^{\bar{q}}\bar{M}_{a}\bar{\zeta}_{a}\right)}\prod_{a}^{q}i^{\frac{1}{2}M_{a}\left(M_{a}-1\right)}Z_{k,M_{a}}^{\left(\mathrm{CS}\right)}\prod_{a}^{\bar{q}}i^{-\frac{1}{2}\bar{M}_{a}\left(\bar{M}_{a}-1\right)}Z_{k,\bar{M}_{a}}^{\left(\mathrm{CS}\right)}\prod_{a,b}^{q,\bar{q}}Z_{M_{a},\bar{M}_{b}}^{\left(\matter\right)}\left(\zeta_{a}-\bar{\zeta}_{b}\right)\nonumber \\
 & \quad\times\prod_{a<b}^{q}\prod_{j_{a},j_{b}}^{M_{a},M_{b}}2\sinh\frac{t_{M_{a},j_{a}}-t_{M_{b},j_{b}}+\zeta_{a}-\zeta_{b}}{2k}\prod_{a<b}^{\bar{q}}\prod_{j_{a},j_{b}}^{\bar{M}_{a},\bar{M}_{b}}2\sinh\frac{t_{\bar{M}_{a},j_{a}}-t_{\bar{M}_{b},j_{b}}+\bar{\zeta}_{a}-\bar{\zeta}_{b}}{2k}\nonumber \\
 & \quad\times e^{-\frac{i\zeta}{k}\left(\sum_{n}^{N_{q+1}}\mu_{n}-\sum_{n}^{\bar{N}_{\bar{q}+1}}\nu_{n}\right)}\prod_{n}^{N_{q+1}}\frac{\prod_{a}^{q}\prod_{j_{a}}^{M_{a}}2\sinh\frac{\mu_{n}+t_{M_{a},j_{a}}+2\pi\zeta_{a}}{2k}}{\prod_{a}^{\bar{q}}\prod_{j_{a}}^{\bar{M}_{a}}2\cosh\frac{\mu_{n}+t_{\bar{M}_{a},j_{a}}+2\pi\bar{\zeta}_{a}}{2k}}\nonumber \\
 & \quad\times\frac{\prod_{m<m'}^{N_{q+1}}2\sinh\frac{\mu_{m}-\mu_{m'}}{2k}\prod_{n<n'}^{\bar{N}_{\bar{q}+1}}2\sinh\frac{\nu_{n}-\nu_{n'}}{2k}}{\prod_{m}^{N_{q+1}}\prod_{n}^{\bar{N}_{\bar{q}+1}}2\cosh\frac{\mu_{m}-\nu_{n}}{2k}}\prod_{n}^{\bar{N}_{\bar{q}+1}}\frac{\prod_{a}^{\bar{q}}\prod_{j_{a}}^{\bar{M}_{a}}2\sinh\frac{\nu_{n}+t_{\bar{M}_{a},j_{a}}+2\pi\bar{\zeta}_{a}}{2k}}{\prod_{a}^{q}\prod_{j_{a}}^{M_{a}}2\cosh\frac{\nu_{n}+t_{M_{a},j_{a}}+2\pi\zeta_{a}}{2k}}.\label{eq:Zcomp4}
\end{align}
Other computation of the substitution into \eqref{eq:Zcomp3} is straightforward. In this step, the phase
\begin{equation}
\prod_{a<b}^{q}i^{M_{a}M_{b}}\prod_{a<b}^{\bar{q}}i^{\bar{M}_{a}\bar{M}_{b}},
\end{equation}
appears in the $\left(2\cosh\right)^{-1}$ factors. This phase can be combined with the phases in \eqref{eq:Zcomp2} and \eqref{eq:Zcomp4} as
\begin{align}
 & i^{\frac{1}{2}\left(N_{q+1}^{2}-N_{1}^{2}+\bar{N}_{1}^{2}-\bar{N}_{\bar{q}+1}^{2}\right)+\frac{1}{2}\left(N_{1}-N_{q+1}-\bar{N}_{1}+\bar{N}_{\bar{q}+1}\right)}\prod_{a}^{q}i^{\frac{1}{2}M_{a}\left(M_{a}-1\right)}\prod_{a}^{\bar{q}}i^{-\frac{1}{2}\bar{M}_{a}\left(\bar{M}_{a}-1\right)}\prod_{a<b}^{q}i^{M_{a}M_{b}}\prod_{a<b}^{\bar{q}}i^{\bar{M}_{a}\bar{M}_{b}}\nonumber \\
 & =i^{-N_{q+1}\left(N_{1}-N_{q+1}\right)+\bar{N}_{\bar{q}+1}\left(\bar{N}_{1}-\bar{N}_{\bar{q}+1}\right)}.\label{eq:phase1}
\end{align}
After the short computation, we finally obtain
\begin{align}
 & i^{\frac{1}{2}\left(N_{q+1}^{2}-N_{1}^{2}+\bar{N}_{1}^{2}-\bar{N}_{\bar{q}+1}^{2}\right)+\frac{1}{2}\left(N_{1}-N_{q+1}-\bar{N}_{1}+\bar{N}_{\bar{q}+1}\right)}\tilde{Z}_{\bm{\zeta},\boldsymbol{N}}^{\left(-q,+1,-\bar{q}\right)}\left(\frac{\mu}{\hbar},\frac{\nu}{\hbar}\right)\nonumber \\
 & =e^{\frac{2\pi i\zeta}{k}\left(\sum_{a}^{q}M_{a}\zeta_{a}-\sum_{a}^{\bar{q}}\bar{M}_{a}\bar{\zeta}_{a}\right)}\prod_{a}^{q}Z_{M_{a}}^{\left(\cs\right)}\prod_{a}^{\bar{q}}Z_{\bar{M}_{a}}^{\left(\cs\right)}\nonumber \\
 & \quad\times\prod_{a<b}^{q}Z_{M_{a},M_{b}}^{\left(\vector\right)}\left(\zeta_{a}-\zeta_{b}\right)\prod_{a,b}^{q,\bar{q}}Z_{M_{a},\bar{M}_{b}}^{\left(\matter\right)}\left(\zeta_{a}-\bar{\zeta}_{b}\right)\prod_{a<b}^{\bar{q}}Z_{\bar{M}_{a},\bar{M}_{b}}^{\left(\vector\right)}\left(\bar{\zeta}_{a}-\bar{\zeta}_{b}\right){\cal Z}_{\bm{\zeta},\boldsymbol{N}}^{q,\bar{q}}\left(\mu,\nu\right).
\end{align}
Therefore, \eqref{eq:FGF-Res} holds.

\subsection{Physical interpretation of prefactors\label{subsec:Factor-Meaning}}

In this appendix, we argue the physical interpretation of $Z_{M,M'}^{\left(\vector\right)}$ and $Z_{M,\bar{M}}^{\left(\matter\right)}$.

First of all, we remember that $Z_{M}^{\left(\cs\right)}$ is the matrix model corresponding to the pure Chern-Simons theory. In the Hanany-Witten setup, the gauge field captures the dynamics of the fundamental strings on the $M$ D3-branes stretched between the NS5-brane and the $\left(1,k\right)$5-brane.
\begin{figure}
\begin{centering}
\includegraphics[scale=0.7]{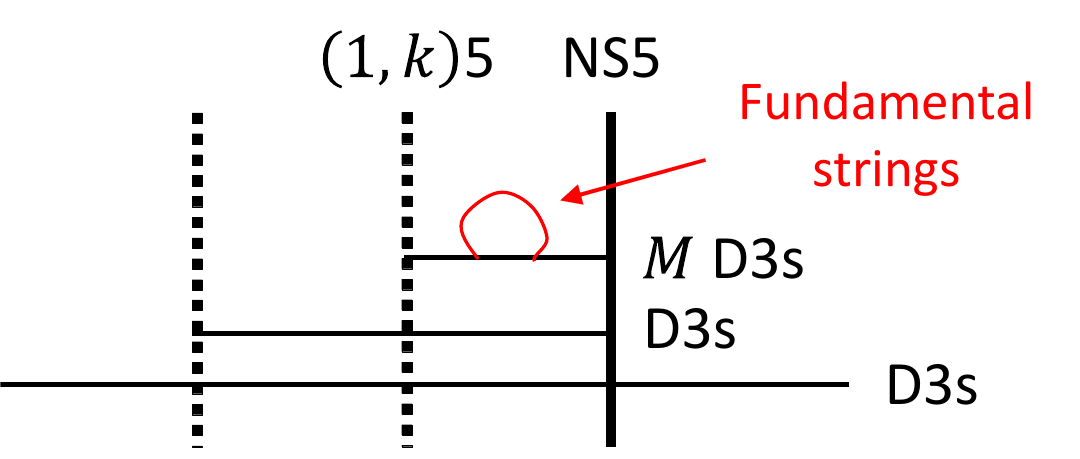}
\par\end{centering}
\caption{The physical interpretations of $Z_{M}^{\left(\protect\cs\right)}$. $Z_{M}^{\left(\protect\cs\right)}$ captures the dynamics of the red fundamental strings.\label{fig:CS-Int}}
\end{figure}
Therefore, $Z_{M}^{\left(\cs\right)}$ captures the dynamics of fundamental strings on the fractional D3-branes (see figure \ref{fig:CS-Int}). Notice that this factor remains when we drop the regular D3-branes because the factor does not depend on the lowest rank. Therefore, this factor does not capture the dynamics of fundamental strings ending on the regular D3-branes.

\begin{figure}
\begin{centering}
\includegraphics[scale=0.7]{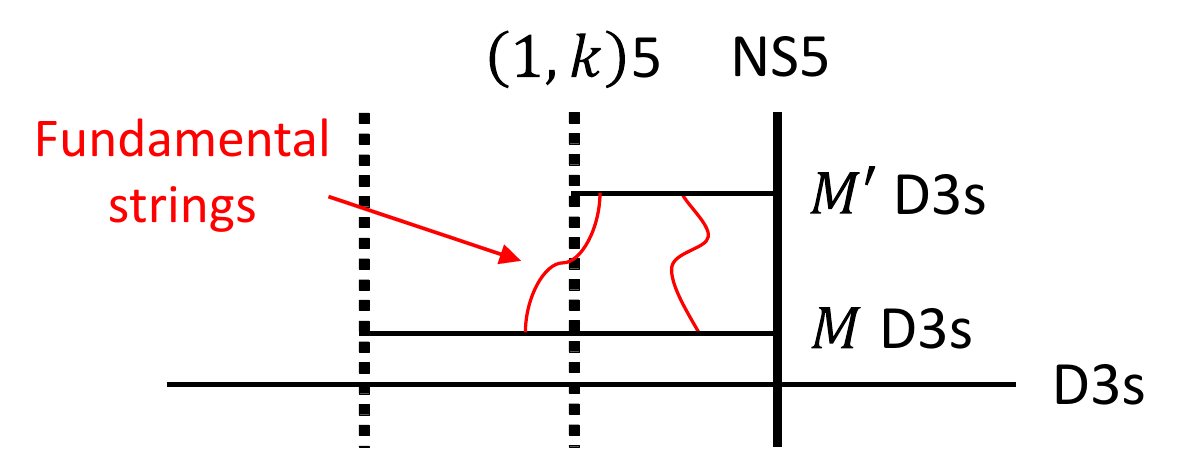}\ \includegraphics[scale=0.7]{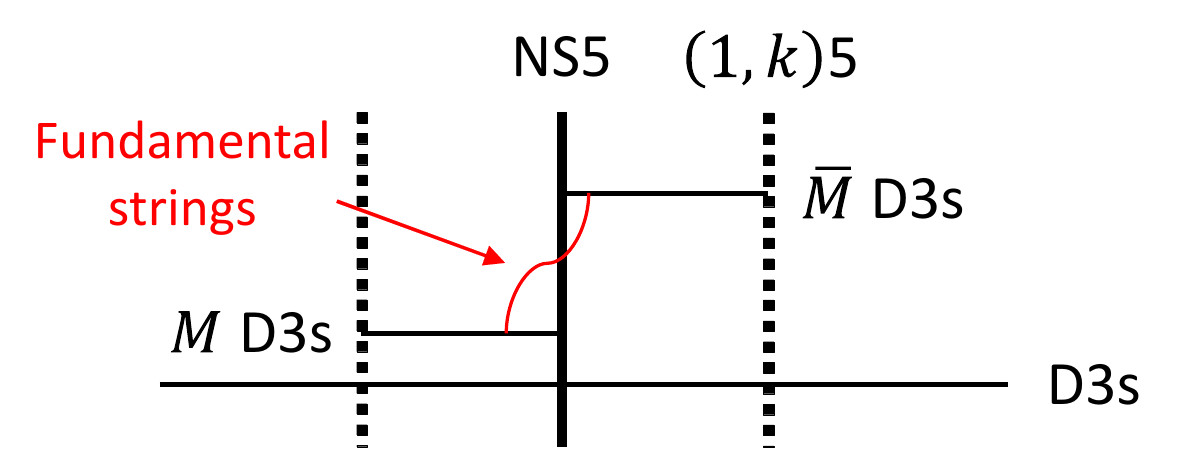}
\par\end{centering}
\caption{The physical interpretations of the prefactors. $Z_{M',M}^{\left(\protect\vector\right)}$ captures the dynamics of the red fundamental strings described in the left side. $Z_{M,\bar{M}}^{\left(\protect\matter\right)}$ captures the dynamics of the red fundamental strings in the right side. \label{fig:vec-mat-Int}}
\end{figure}

It is worth noting that $Z_{M,M'}^{\left(\vector\right)}$ and $Z_{M,\bar{M}}^{\left(\matter\right)}$ also remain when we drop the regular D3-branes. We claim that $Z_{M,M'}^{\left(\vector\right)}$ and $Z_{M,\bar{M}}^{\left(\matter\right)}$ capture the dynamics of fundamental strings between fractional D3-branes as shown in figure \ref{fig:vec-mat-Int}.
In the following, we give pieces of evidence.

In the Fermi gas computation, $Z_{M',M}^{\left(\vector\right)}$ comes from $\sinh$ factors in the numerator of the NS5-factor $Z_{\zeta,\left(N_{1},\bar{N}_{1}\right)}^{\left(+1\right)}$ in \eqref{eq:Zcomp4}. Since this factor comes from 1-loop determinant of the gauge field, it captures the dynamics of fundamental strings stretched between the $M$ D3-branes and the $M'$ D3-branes (the right fundamental strings at the first brane configuration in figure \ref{fig:vec-mat-Int}). There is additional contribution for $Z_{M',M}^{\left(\vector\right)}$ from $\left(\cosh\right)^{-1}$ factors in \eqref{eq:Zcomp3}. Since these factors come from the $\left(1,k\right)$5-brane factor $Z_{\zeta_{a},\left(N_{a},N_{a+1}\right)}^{\left(-1\right)}$, these factors capture the dynamics of fundamental strings which are stretched between the $M$ D3-branes and the $M'$ D3-branes and cross the $\left(1,k\right)$5-brane (the left fundamental strings at the first brane configuration in figure \ref{fig:vec-mat-Int}). In summary, $Z_{M',M}^{\left(\vector\right)}$ captures the dynamics of two types of fundamental strings at the first brane configuration in figure \ref{fig:vec-mat-Int}. On the other hand, $Z_{M,\bar{M}}^{\left(\matter\right)}$ comes from $\cosh$ factors in the denominator of the NS5-brane factor $Z_{\zeta,\left(N_{1},\bar{N}_{1}\right)}^{\left(+1\right)}$ in \eqref{eq:Zcomp4}. Since these factors come from the 1-loop determinant of the bi-fundamental matter, these factors capture the dynamics of fundamental strings stretched between the $M$ D3-branes and the $\bar{M}$ D3-branes (the fundamental strings at the second brane configuration in figure \ref{fig:vec-mat-Int}).

As another evidence, the difference of the FI parameters denotes the difference of the positions of the 5-branes to the transverse directions. The argument of $Z_{M,M'}^{\left(\vector\right)}$ and $Z_{M,\bar{M}}^{\left(\matter\right)}$ are the difference of the FI parameters. This is consistent with figure \ref{fig:vec-mat-Int}.

The first and the second brane configurations in figure \ref{fig:vec-mat-Int} are related by the Hanany-Witten transition, where the relation of the parameters is $\bar{M}=k-M'$. We can show that \cite{Kubo:2020qed}
\begin{equation}
Z_{M',M}^{\left(\vector\right)}=Z_{k-M',M}^{\left(\matter\right)}.
\end{equation}
Under the Hanany-Witten transition, the fundamental strings at the first and the second brane configurations in figure \ref{fig:vec-mat-Int} correspond to each other. Therefore, this is the other evidence of our claim.

\section{Hanany-Witten transition and matrix model\label{sec:HW}}

When an NS5-brane and a $\left(1,k\right)$5-brane pass through each other, the $k$ D3-branes are created \cite{Hanany:1996ie}. This transition is called the Hanany-Witten transition. Therefore, the Hanany-Witten transition changes the number of D3-branes as
\begin{equation}
\left\langle N_{{\rm L}}\bullet_{\zeta}N\circ_{\zeta'}N_{{\rm R}}\right\rangle \sim\left\langle N_{{\rm L}}\circ_{\zeta'}\tilde{N}\bullet_{\zeta}N_{{\rm R}}\right\rangle ,\label{eq:HW-Def}
\end{equation}
where $\tilde{N}=N_{{\rm L}}+N_{{\rm R}}-N+k$. The matrix model enjoys the duality generated by the Hanany-Witten transition \cite{Assel:2014awa}. The Fermi gas formalism is useful to show the identity. In this section, we shortly comment on it.

The local matrix model corresponding to the left side of \eqref{eq:HW-Def} in the Fermi gas formalism is \eqref{eq:FGF-Res}:
\begin{align}
Z_{\bm{\zeta},\boldsymbol{N}}^{\left(+1,-1\right)}\left(\frac{\alpha}{\hbar},\frac{\beta}{\hbar}\right)= & e^{i\Theta_{\bm{\zeta},\boldsymbol{N}}^{0,1}}e^{-\frac{2\pi i}{k}\zeta\zeta'M}Z_{M}^{\left(\cs\right)}\int d^{N_{{\rm L}}}\lambda d^{N_{{\rm R}}}\lambda'\nonumber \\
 & \times\prod_{n}^{N_{{\rm L}}}\braket{\alpha_{n}|e^{\frac{i}{2\hbar}\hat{p}^{2}}|\lambda_{n}}{\cal Z}_{\bm{\zeta},\boldsymbol{N}}^{0,1}\left(\lambda,\lambda'\right)\prod_{n}^{N_{{\rm R}}}\braket{\lambda_{n}'|e^{-\frac{i}{2\hbar}\hat{p}^{2}}|\beta_{n}},
\end{align}
where $\bm{\zeta}=\left(\zeta,\zeta'\right)$, $\boldsymbol{N}=\left(N_{{\rm L}},N,N_{{\rm R}}\right)$ and $M=N-N_{{\rm R}}$. The local matrix model corresponding to the right side of \eqref{eq:HW-Def} is
\begin{align}
Z_{\bm{\zeta}',\boldsymbol{N}'}^{\left(-1,+1\right)}\left(\frac{\alpha}{\hbar},\frac{\beta}{\hbar}\right)= & e^{i\Theta_{\bm{\zeta}',\boldsymbol{N}'}^{1,0}}e^{\frac{2\pi i}{k}\zeta\zeta'\left(k-M\right)}Z_{k-M}^{\left(\cs\right)}\int d^{N_{{\rm L}}}\lambda d^{N_{{\rm R}}}\lambda'\nonumber \\
 & \times\prod_{n}^{N_{{\rm L}}}\braket{\alpha_{n}|e^{\frac{i}{2\hbar}\hat{p}^{2}}|\lambda_{n}}{\cal Z}_{\bm{\zeta}',\boldsymbol{N}'}^{1,0}\left(\lambda,\lambda'\right)\prod_{n}^{N_{{\rm R}}}\braket{\lambda_{n}'|e^{-\frac{i}{2\hbar}\hat{p}^{2}}|\beta_{n}},
\end{align}
where $\bm{\zeta}'=\left(\zeta',\zeta\right)$ and $\boldsymbol{N}'=\left(N_{{\rm L}},\tilde{N},N_{{\rm R}}\right)$. We ignore the phase factor since they are the framing factor. The equivalence of $Z_{M}^{\left(\cs\right)}$ and $Z_{k-M}^{\left(\cs\right)}$ is known \cite{Kapustin:2010mh}. We can also easily show the equivalence between ${\cal Z}_{\bm{\zeta},\boldsymbol{N}}^{0,1}$ and ${\cal Z}_{\bm{\zeta}',\boldsymbol{N}'}^{1,0}$. From the definition \eqref{eq:zFermi-Def}, we find
\begin{align}
{\cal Z}_{\bm{\zeta},\boldsymbol{N}}^{0,1}\left(\lambda,\lambda'\right)= & C_{\zeta',M}\left(\mu\right)S_{\zeta',M}\left(\nu\right)e^{-\frac{i\zeta}{k}\left(\sum_{m}^{N_{{\rm L}}}\mu_{m}-\sum_{n}^{N_{{\rm R}}}\nu_{n}\right)}\nonumber \\
 & \times\frac{\prod_{m<m'}^{N_{{\rm L}}}2\sinh\frac{\lambda_{m}-\lambda_{m'}}{2k}\prod_{n<n'}^{N_{{\rm R}}}2\sinh\frac{\lambda_{n}'-\lambda_{n'}'}{2k}}{\prod_{m}^{N_{{\rm L}}}\prod_{n}^{N_{{\rm R}}}2\cosh\frac{\lambda_{m}-\lambda_{n}'}{2k}},
\end{align}
and
\begin{align}
{\cal Z}_{\bm{\zeta}',\boldsymbol{N}'}^{1,0}\left(\lambda,\lambda'\right)= & S_{\zeta',k-M}\left(\mu\right)C_{\zeta',k-M}\left(\nu\right)e^{-\frac{i\zeta}{k}\left(\sum_{m}^{N_{{\rm L}}}\mu_{m}-\sum_{n}^{N_{{\rm R}}}\nu_{n}\right)}\nonumber \\
 & \times\frac{\prod_{m<m'}^{N_{{\rm L}}}2\sinh\frac{\lambda_{m}-\lambda_{m'}}{2k}\prod_{n<n'}^{N_{{\rm R}}}2\sinh\frac{\lambda_{n}'-\lambda_{n'}'}{2k}}{\prod_{m}^{N_{{\rm L}}}\prod_{n}^{N_{{\rm R}}}2\cosh\frac{\lambda_{m}-\lambda_{n}'}{2k}}.
\end{align}
By using the identity \cite{Kubo:2020qed}
\begin{equation}
\frac{\prod_{j}^{M}2\sinh\frac{x+t_{M,j}}{2k}}{e^{\frac{x}{2}}+\left(-1\right)^{M}e^{-\frac{x}{2}}}=\frac{1}{\prod_{j}^{k-M}2\cosh\frac{x+t_{k-M,j}}{2k}},
\end{equation}
we find
\begin{equation}
S_{\zeta,M}\left(x\right)=C_{\zeta,k-M}\left(x\right).
\end{equation}
This immediately leads to ${\cal Z}_{\bm{\zeta},\boldsymbol{N}}^{0,1}={\cal Z}_{\bm{\zeta}',\boldsymbol{N}'}^{1,0}$.

\printbibliography

\end{document}